\documentclass[12pt]{article}

\usepackage[latin1]{inputenc}
\usepackage{amsmath}
\usepackage{amsfonts}
\usepackage{amssymb,MnSymbol,slashed}
\usepackage{amsxtra}
\usepackage[margin=3cm]{geometry}
\usepackage{color}
\usepackage{hyperref}
\hypersetup{linktocpage=true}
\usepackage{graphicx}
\usepackage{placeins}
\usepackage{upgreek} 
\usepackage{mathrsfs}
\usepackage{accents}
\usepackage{multirow} 
\usepackage{tikz}

\numberwithin{equation}{section}
\newcommand{\beq}{\begin{equation}}
\newcommand{\eeq}{\end{equation}}
\def\be {\begin{equation}}
\def\ee {\end{equation}}
\def\bs#1\es{\begin{split}#1\end{split}}
\def\ba#1\ea{\begin{align}#1\end{align}}
\def\bg#1\eg{\begin{gathered}#1\end{gathered}}
\def\bea{\begin{eqnarray}}
\def\eea{\end{eqnarray}}

\def\nn{\nonumber}

\def\a{\alpha}
\def\b{\beta}
\def\c{\chi}
\def\d{\delta}

\def\e{\epsilon}

\def\f{\phi}
\def\vf{\varphi}
\def\F{\Phi}
\def\g{\gamma}
\def\G{\Gamma}
\def\h{\eta}

\def\l{\lambda}
\def\L{\Lambda}
\def\m{\mu}
\def\n{\nu}
\def\o{\omega}
\def\O{\Omega}
\def\p{\psi}

\def\Q{\Theta}
\def\r{\rho}
\def\s{\sigma}
\def\S{\Sigma}

\def\x{\xi}
\def\z{\zeta}

\def\bls{\bigg [}
\def\brs{\bigg ]}

\def\R{\text{Re}}

\newcommand{\cT}{\mathcal{T}}

\newcommand{\cC}{\mathcal{C}}

\newcommand{\cK}{\mathcal{K}}

\newcommand{\cG}{\mathcal{G}}

\newcommand{\cB}{\mathcal{B}}
\newcommand{\cF}{\mathcal{F}}
\newcommand{\cI}{\mathcal{I}}

\def\cD{\mathcal{D}}
\def\cL{\mathcal{L}}

\def\cA{{{\mathcal A}}}
\def\cO{{{\mathcal O}}}
\def\cM{\mathcal{M}} 
\def\cN{\mathcal{N}}
\def\cV{\mathcal{V}}

\def\bP{\mathbb{P}}
\def\bZ{\mathbb{Z}}
\def\bR{\mathbb{R}}

\def\Re{\text{Re}}
\def\Im{\text{Im}}
\def\Tr{\text{Tr}}

\def\sq{\sqrt}
\def\pa{\partial}
\def\na{\nabla}
\def\fr{\frac}
\def\tfr{\tfrac}
\def\id{\rlap 1\mkern4mu{\rm l}}
\def\we{\wedge}
\def\ra{\rightarrow}

\newcommand{\wh}[1]{ {\hat{#1}}{} }
\newcommand{\til}[1]{ {\tilde{#1}} }

\let\foo\bar 
\renewcommand{\bar}[1]{ {\foo{  #1} }{} }

\newlength{\dhatheight}

\def\Vz{ \hat \cV}
\def\Vy{ \cV }
\def\Vn{N}
\def\Jn{{J_\f}}
\def\vn{\f}

\def\resZ{{\hat Z}}

\begin{document}

\baselineskip=16pt
\setlength{\parskip}{6pt}

\begin{titlepage}
\begin{flushright}
\parbox[t]{1.4in}{
\flushright MPP-2013-194}
\end{flushright}

\begin{center}

\vspace*{1.5cm}

{\Large \bf  Non-Supersymmetric F-Theory Compactifications\\[.2cm]
 on Spin(7) Manifolds}

\vskip 1.5cm

\renewcommand{\thefootnote}{}

\begin{center}
 \normalsize \bf{Federico Bonetti, Thomas W.~Grimm, Tom G.~Pugh}\footnote{\texttt{bonetti,\ grimm,\ pught @mpp.mpg.de}} 
\end{center}
\vskip 0.5cm

 \emph{ Max Planck Institute for Physics, \\ 
        F\"ohringer Ring 6, 80805 Munich, Germany} 
\\[0.25cm]

\end{center}

\vskip 1.5cm
\renewcommand{\thefootnote}{\arabic{footnote}}

\begin{center} {\bf ABSTRACT } \end{center}
We propose a novel approach to obtain 
non-supersymmetric four-dimensional effective actions 
by considering F-theory on manifolds with special holonomy Spin(7).
To perform such studies we suggest that a duality
relating M-theory on a certain class of Spin(7) manifolds 
with F-theory on the same manifolds times an interval exists. 
The Spin(7) geometries under consideration are constructed 
as quotients of elliptically fibered Calabi-Yau fourfolds 
by an anti-holomorphic and isometric involution. 
The three-dimensional minimally supersymmetric
effective action of M-theory on a general Spin(7) manifold 
with fluxes is determined and specialized to the aforementioned 
geometries. This effective theory is compared 
with an interval Kaluza-Klein reduction of a non-supersymmetric 
four-dimensional theory with definite boundary conditions for all fields.
Using this strategy a minimal set of couplings of the 
four-dimensional low-energy effective actions
is obtained in terms of the Spin(7) geometric data.
We also discuss briefly the string interpretation in the Type IIB weak coupling limit.

\end{titlepage}

\newpage
\noindent\rule{15cm}{.1pt}		
\tableofcontents
\vspace{20pt}
\noindent\rule{15cm}{.1pt}

\setcounter{page}{1}
\setlength{\parskip}{9pt}

\newpage

\section{Introduction}
Over the last decades four-dimensional (4d) supersymmetric effective theories arising in string 
compactifications have been studied intensively. Minimally supersymmetric theories 
are considered as providing interesting physics beyond the Standard Model.
Therefore it has been a crucial long-standing task to 
embed supersymmetric extensions of the Standard Model or Grand Unified Theories 
into string theory as reviewed, for example, in \cite{Blumenhagen:2005mu, Blumenhagen:2006ci,
Nilles:2008gq,Weigand:2010wm,Maharana:2012tu}. The established approach is to consider 
compactifications of string theory on manifolds with special holonomy, such 
that some of the underlying ten-dimensional (10d) supersymmetries are preserved 
in four dimensions and allow a supersymmetric effective theory to be determined.  
Precisely these supersymmetry-preserving geometries are also mathematically best studied
and many powerful tools have been developed exploiting the interplay of 
geometry and low-energy physics. In this work we will examine whether one can find a 
rich set of string compactifications with non-supersymmetric 4d 
effective theories, and possibly interesting phenomenological properties, while still allowing the 
virtues of the remarkable
 mathematical tools developed for special holonomy manifolds to be used.

Our considerations are based on the study of F-theory compactifications
to four dimensions. Recall that F-theory vacua describe the geometry 
of Type IIB string compactifications with varying complexified 
string coupling constant. This change of coupling is encoded by the complex 
structure of an auxiliary two-torus, which varies over the ten-dimensional 
space-time of the Type IIB theory. Vacua of F-theory are thus torus 
fibrations over some base space that provides the hidden compact 
dimensions of Type IIB string theory. This implies that F-theory 
compactifications to four space-time dimensions require an eight-dimensional 
compact and torus-fibered geometry to be specified. Furthermore,
singularities of this fibration indicate the presence of space-time filling 
seven-branes. Therefore, this setup geometrizes many aspects 
of open string physics and hence allows the construction of many interesting phenomenological 
models. 

Minimal supersymmetry, for which the 4d effective theory 
has four real supercharges, is preserved by the geometry if the 
compact eight-dimensional space 
has  SU(4)  holonomy, i.e.~is a Calabi-Yau fourfold \cite{Vafa:1996xn, Denef:2008wq}. However, 
on eight-dimensional manifolds the classification by Berger \cite{berger} shows 
that  SU(4)  is \textit{not} the maximal 
possible special holonomy group within the local Lorentz group SO(8). 
This maximal special holonomy group is instead given by Spin(7).
In what follows we will refer to these manifolds with 
Spin(7) holonomy as simply \textit{Spin(7) manifolds}.
For these geometries one therefore is 
led to ask: 
\vspace*{-.2cm}
\begin{itemize}
 \item[(1)] Is there a controlled construction of Spin(7) manifolds that 
 can serve as backgrounds for F-theory?

 \item[(2)] What are the characteristics of the 4d non-supersymmetric effective theories 
arising from F-theory compactifications on such Spin(7) manifolds?

 \item[(3)] What is the weak coupling Type IIB string interpretation of these theories?
\end{itemize}
In this work we will attempt to systematically address these questions. 
It should be noted that the consideration of F-theory 
on Spin(7) manifolds was already mentioned in the original paper by Vafa \cite{Vafa:1996xn}, in connection 
with the proposals of Witten \cite{Witten:1994cga,Witten:1995rz}. However, this link has not be concretized since then. 
With the recent progress on deriving the 4d supersymmetric effective 
action of F-theory on Calabi-Yau fourfolds \cite{Grimm:2010ks}, we are now endowed with the 
necessary advances to suggest  a concrete F-theory 
and string theory construction.

Before even entering any analysis of the effective action, we have to answer 
the question of whether or not there are suitable Spin(7) manifolds that can be used for F-theory. 
In particular, it will be crucial to single out geometries that have an appropriate torus 
fibration structure to identify the F-theory compactification
as a Type IIB string background. In building these manifolds we will be 
motivated by the constructions described by Joyce \cite{Joyce:1999nk}. These 
constructions begin by considering a Calabi-Yau fourfold which is then 
quotiented in such a way that a Spin(7) manifold is generated. Here we will 
investigate whether this process, carried out for   elliptically fibered 
Calabi-Yau fourfolds, may generate appropriate Spin(7) manifolds for 
use in these F-theory compactifications. It should be stressed that one expects 
that there exist many more examples 
of Spin(7) geometries that are not based on any Calabi-Yau fourfold. 
Definite statements about these more general cases turn out to be hard 
to extract, nevertheless various results of our analysis may well extend beyond the 
context that we consider. Importantly, these constructions based 
upon Calabi-Yau quotients give us control
over the setup and allow our intuition about Calabi-Yau fourfold compactifications 
of F-theory to be used. Other explicit constructions of Spin(7) geometries appeared in \cite{Cvetic:2001pga,Cvetic:2001zx}.

To derive effective physics of these F-theory compactifications it will be necessary 
to take a detour via M-theory. This can be traced back to the fact
that there is no fundamental low-energy effective action of F-theory. M-theory 
has eleven-dimensional (11d) supergravity as a low-energy effective action \cite{Cremmer:1978km}
and hence provides a well-defined setup to study compactifications on 
smooth compact geometries. In fact, if one considers M-theory 
on a Spin(7) geometry one obtains a three-dimensional (3d) effective 
theory with minimal supersymmetry, i.e.~two supercharges \cite{Papadopoulos:1995da}.\footnote{These
compactifications are in fact on warped backgrounds, but we will not consider the impact
of warping in this work.} 
We determine the 3d effective action of M-theory on a general Spin(7) manifold with probe
fluxes extending and applying earlier works \cite{Becker:2000jc,Gukov:2001hf,Curio:2001dz,Acharya:2002vs,Gukov:2002zg,Becker:2003wb} and determine the 
couplings in terms of the geometric data of the Spin(7) geometry. 
To take the F-theory limit of this 3d
theory to four space-time dimensions we propose the following 
duality:
\beq \label{new_dual}
   \textit{M-theory on Spin(7) manifold}  \ \cong \  \textit{F-theory on}\ \left\{\begin{array}{c}
                                                            \textit{Spin(7) manifold} \\
                                                             \! \!  \textit{(with vanishing fiber)}
                                                            \end{array}   \times \  \textit{Interval}\ \right\}
\ .
\eeq

To provide evidence for  \eqref{new_dual} we consider a certain 
non-supersymmetric 4d theory on an interval. If the interval is small, 
we perform a Kaluza-Klein reduction to an effective 3d theory. Specifying 
the definite boundary conditions for the various fields, we argue that 
the 3d effective theory of the zero modes is minimally supersymmetric 
and can be identified with the effective theory arising from 
a compactification of M-theory on a Spin(7) manifold. 
The original 4d theory should be recovered in the limit in which the 
interval length is sent to infinity. This should correspond to 
sending the fiber volume of the Spin(7) manifold to zero and 
provide a realization of the M-theory to F-theory limit. However,
in this work we will mostly consider a finite size interval either in the 
derivation of the 3d effective theory, or in the 4d lift to an effectively 
non-supersymmetric theory due to boundary effects. 

One difference to the M-theory to F-theory limit for Calabi-Yau fourfolds is
the appearance of an interval instead of a circle. This interval is 
crucial as the boundary conditions that are imposed project out half of the zero mode 
degrees of freedom that would arise  in the circle reduction 
of a 4d fermion. 
This means that on the level of 3d zero modes only a part of the 
4d fermionic degrees of freedom have to be completed with bosonic counterparts. 
This allows a non-supersymmetric 
spectrum in four dimensions to be dimensionally reduced to a minimally supersymmetric zero mode 
spectrum in three dimensions. The appearance of an interval is also natural from the construction 
of Spin(7) manifolds that we have mentioned above for which the quotient 
of the fourfold may be associated with the quotient of the circle that gives rise to the interval. 
It is crucial in \eqref{new_dual} that the core features of the non-supersymmetric theory in four dimensions 
and the boundary conditions for the interval are fixed by the Spin(7) geometry. 

\begin{figure}[h]
 \centering
\begin{minipage}{0.9\textwidth}
 \centering
\includegraphics[width=0.8\textwidth]{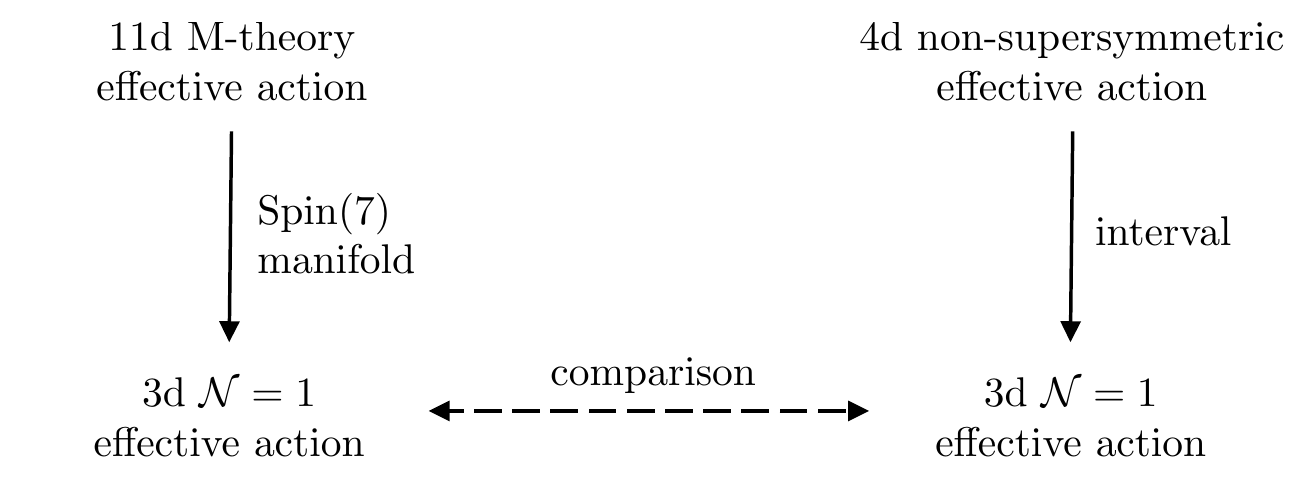}
\caption{\small
Summary of the effective actions considered in this work. The left column corresponds to the M-theory side of the
duality \eqref{new_dual}, while the right column corresponds to the F-theory side.
The comparison between the 3d $\mathcal N=1$
theories is performed in the case in which the Spin(7) manifold arises as an 
anti-holomorphic quotient of an elliptically fibered Calabi-Yau fourfold.
We consider a fibration structure that yields a simple non-Abelian gauge group.
The match of 3d actions is carried out in the Coulomb branch at the level of zero modes.
}
\label{fig:summary}
\end{minipage}
\end{figure}

In this work we will provide evidence for \eqref{new_dual} in the context of the above mentioned  
quotiented Calabi-Yau geometries, and discuss important parts of the 3d and 4d 
effective actions of M-theory and F-theory. 
A schematic picture of the effective  actions considered in the
following sections can be found in figure \ref{fig:summary}.
We do believe, however, that such an analysis should 
similarly be possible for other classes of Spin(7) manifolds with an 
appropriate fibration structure.

\section{Geometries with Spin(7) Holonomy for F-theory}
\label{Geoms}

To set the stage for the discussions that follow we first recall some facts about 
Spin(7) manifolds and their construction. In subsection \ref{constructingSpin7}
we give a brief introduction to aspects of the differential and algebraic geometry 
of Spin(7) manifolds.  We also describe the 
construction of Spin(7) manifolds as anti-holomorphic quotients 
of Calabi-Yau fourfolds. This construction is applied to 
elliptically fibered Calabi-Yau fourfolds in subsection \ref{sec:spin7fromCY}. We discuss 
the fiber structures which arise and comment on 
seven-brane configurations that can appear.

\subsection{Constructing Spin(7) Manifolds from Calabi-Yau Fourfolds} \label{constructingSpin7}

Let us briefly recall certain important features of the geometry of Spin(7) holonomy eight-dimensional 
manifolds, which we will refer to as \textit{Spin(7) manifolds}. 
To do this it is convenient to begin by analyzing the set of independent covariantly constant 
spinors that may exist on such a space $Z_8$. All spinors on $Z_8$ will transform as definite 
representations of the holonomy group and so their properties may be studied by decomposing 
the representations of SO(8) (the holonomy group of an orientable eight-dimensional Riemannian
manifold) under Spin(7). In doing this we find that the representations corresponding to Majorana-Weyl 
spinors decompose as
\ba
\bold{8}_+ &\ra \bold{1} \oplus \bold{7}\, ,  & 
&\text{and}& 
\bold{8}_- &\ra \bold{8} \, . 
\ea
The singlets present in this decomposition determine the covariantly constant spinors of $Z_8$ so a given 
Spin(7) manifold may have only one independent covariantly constant spinor which we will call $\eta$. 
From this spinor we may construct the covariantly constant nowhere-vanishing $p$-forms of $Z_8$ by 
taking contractions with the gamma matrices in the usual way. However as $\eta$ is Majorana-Weyl with 
positive chirality the only non trivial 
$p$-form that may be constructed is a self-dual four-form
\ba \label{eq:Phi_expr}
\F_{m n r s} &= \bar \eta \, \g_{m n r s} \, \eta \, ,& &\text{where}&
\fr{1}{\Vz} \int_{Z_8} \F \we \F & =  || \F ||^2 = \fr{1}{ 4!} \F_{m n r s} \F^{m n r s}\, ,
\ea
and where $\Vz$ is the volume of $Z_8$. This four-form then gives the Cayley calibration of $Z_8$.
We note here that by using Fierz identities one may show that $\F$ satisfies the useful identity
\ba \label{eq:useful}
\F^{m n p t} \F_{q r s t } &=  \fr{3}{7} ||\F||^2  \d_{[ q}^m \d_r^n \d_{s]}^p - \fr{9}{\sqrt{14}}  ||\F|| \d_{[q}{}^{[m} \F_{rs]}{}^{np]} \, . 
\ea
In a similar way one may analyze the cohomology of the Spin(7) manifold by decomposing the 
various cohomology groups under Spin(7). This then gives \cite{JoyceBook}
\ba
H^0 (Z_8,\bR) & = \bR \, ,& 
H^1 (Z_8,\bR) & = 0 \, , & 
H^2 (Z_8,\bR) & = H^2_{\bold{21}}(Z_8,\bR) \, , &  \nn 
\ea
\vspace{-1.2cm}
\ba
H^3 (Z_8,\bR) & = H^3_{\bold{48}}(Z_8,\bR)  \, , & 
H^4 (Z_8,\bR) & = H^4_{\bold{1} \, \text{S}}(Z_8,\bR) \oplus H^4_{\bold{27} \, \text{S} }(Z_8,\bR) \oplus H^4_{\bold{35} \, \text{A}}(Z_8,\bR) \, , 
\ea
where S and A indicate the self-duality and anti-self-duality of the four-forms respectively. The only in-equivalent representative of  $H^4_{\bold{1} \, \text{S}}(Z_8,\bR)$ is then given by $\F$. 
The Betti numbers $b^{n}(Z_8) =\text{dim}(H^n (Z_8,\bR) )$ satisfy one constraint,
\ba
b^2(Z_8) - b^3(Z_8) - b^4_S(Z_8) +2 b^4_A(Z_8) + 25 = 0 \ .
\ea
This implies that there are three independent Betti numbers, for example, $b^{2}(Z_8)$, $b^{3}(Z_8)$ and $b^{4}_{A}(Z_8)$.

By contrast a Calabi-Yau fourfold $Y_4$ has both a covariantly constant $(1,1)$ 
K\"ahler form $J$ and a holomorphic $(4,0)$ 
form $\Omega$. In \cite{Joyce:1999nk} these are related to a self-dual four-form $\F$ 
by considering an anti-holomorphic and isometric involution $\sigma:Y_4 \rightarrow Y_4$, i.e.~$\sigma$ satisfies
 \ba
    \sigma^2 &= \id \, , &   
     & \left\{\begin{array}{ll}
                                   \text{isometric} & \sigma^*(g) = g\ , \\
                                   \text{anti-holomorphic} & \sigma^*(I) = - I\ ,
                                  \end{array} \right. 
 \ea
 where $g$ and $I$ are the metric and complex structure on $Y_4$, 
 respectively.
 These conditions translate to the forms $J$ and $\Omega$ as 
 \ba \label{eq:OmegaJtransform}
    \sigma^* J &=  - J\ , & 
    \sigma^* \Omega &= e^{2i\theta} \bar \Omega\ ,
 \ea
 where $\theta$ is some constant phase factor. The forms $J$ and $\Omega$ then naturally define a Spin(7)-structure on $Y_4$ with $\Phi$ given by 
\ba \label{eq:Phi_preliminary}
\Phi &= \fr{1}{\Vy^2}(\fr{1}{||\O||}\text{Re} (e^{-i\theta} \Omega) + \frac{1}{8 } J \wedge J)  \, , &&\text{where} & \Vy &= \frac{1}{4!} \int_{Y_4} J^4 \, , 
\ea
is the volume of $Y_4$ and $||\O||$ is defined analogously to \eqref{eq:Phi_expr}. The derivation 
of the precise prefactors in front of $\text{Re} (e^{-i\theta} \Omega) $ and $J \wedge J$ will be presented 
in section \ref{sec:MonSpin7fromCY}.
The four-form $\Phi$ 
is invariant under the involution $\sigma$ and an associated Spin(7) manifold may then be constructed by quotienting $Y_4$ by $\sigma$ and resolving the singularities 
in a Spin(7) compatible way \cite{Joyce:1999nk}. In this way $Y_4$ represents the double cover of $Z_8$ which relates the volumes as $\Vy = 2 \Vz$.

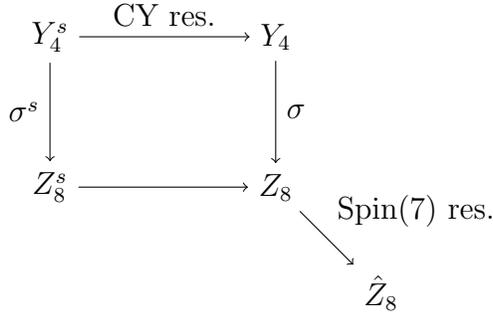
\begin{figure}
 \begin{center}
 \begin{minipage}{0.8\textwidth}
 \begin{center}
\begin{tikzpicture}[node distance=2cm, auto]
  \node (Y) {$Y^s_4$};
  \node (Yh) [node distance=3cm, right of=Y] {$ Y_4$};
  \node (Zt) [node distance=2cm, below of=Y] {$Z^s_8 $};
  \node (Z) [node distance=2cm, below of=Yh] {$Z_8$};
  \node (Zh) [node distance=1.4cm,  right of=Z, below of=Z] {$\hat{Z}_8$};
  \draw[->] (Y) to node {CY\ res.} (Yh);
  \draw[->] (Y) to node [swap] {$\s^s$} (Zt);
  \draw[->] (Zt) to node [swap] {} (Z);
  \draw[->] (Yh) to node {$\s$ } (Z);
  \draw[->] (Z) to node {Spin(7) res.} (Zh);
\end{tikzpicture}
\caption{Construction of Spin(7) manifolds by using Calabi-Yau fourfolds with anti-holomorphic involutions.}
\label{fig:resolution_spaces}
\end{center}
\end{minipage}
\end{center}
\end{figure}

In preparation for the application to F-theory let us comment further on the involved geometries.
We note that when considering F-theory on a Calabi-Yau space $Y_4^s$, the space can be chosen to be
singular. The singularities arise, for example, when the 4d 
theory has to have a non-Abelian gauge group. These non-Abelian singularities can 
be resolved in a way that is compatible with the Calabi-Yau condition to yield a manifold 
$Y_4$. We denote the anti-holomorphic involution on the singular space $Y_4^s$ by $\sigma^s$ 
and on the resolved space by $\sigma$. The respective quotient spaces 
are denoted by $Z^s_8 = Y_4^s / \sigma^s$ and $Z_8 = Y_4/ \sigma$.
The Spin(7) resolution of $Z_8$ will be denoted by $\hat Z_8$.
By analogy with the standard M-theory/F-theory duality
we thus expect that  
 the duality  \eqref{new_dual} relates F-theory compactified on $Z^s_8$ with M-theory compactified on $\hat Z_8$.  
It should be stressed that finding a resolution 
of $Z_8$ admitting a Spin(7) structure is a hard task and 
involves constructing local real Spin(7) ALE geometries that 
can be used to resolve possible orbifold singularities \cite{Joyce:1999nk}.
The Betti numbers of the resolved space can be computed as described in 
\cite{Joyce:1999nk}. A stringy computation of the Betti numbers on the 
quotient geometry $Z_8$ can be found 
in \cite{Blumenhagen:2001qx}.
In this work we will not be concerned with this real resolution $\hat Z_8$, and
mostly work with $Z_8$ neglecting possible singularities. We 
will refer to the Spin(7) manifold $Z_8$ constructed in this way 
as a \textit{quotient torus fibration}. Our goal 
is, however, to formulate the results in a general Spin(7) language such 
that they can be equally applied to the resolved geometries $\hat Z_8$.
We summarize the relevant geometries in figure  \ref{fig:resolution_spaces}.

The construction that is carried out in \cite{Joyce:1999nk} assumes 
certain additional properties of the orbifold singularities that are 
required for the Spin(7) ALE resolutions which are considered there 
to be applied. One such condition is that the singularities introduced 
by quotienting with respect to $\s$ must be isolated points in $Z_8$ 
which lie at points that are already holomorphic orbifold singularities of $Y_4$. 
However it is anticipated that these resolution methods are by no means the 
only possibility. Therefore, in what follows, we will not limit ourselves to 
considering only the sorts of singularities which are required in \cite{Joyce:1999nk}, 
but will bear in mind these additional constraints. The analysis of the more general 
resolutions that would then be required and the physics associated with their 
structure will not be discussed here and therefore represents an important 
topic for consideration in future work.

\subsection{Spin(7) Manifolds from Calabi-Yau Elliptic Fibrations}
\label{sec:spin7fromCY}

In order that the Spin(7) manifold $Z_8$ can be used as a background 
of F-theory we require that the Calabi-Yau fourfold $Y_4$ is 
an elliptic fibration with K\"ahler base $B_3$. The elliptic 
fiber $\cC_p$ at a point $p$ on $B_3$ 
can always be described by a Weierstrass equation \footnote{The precise statement is that 
every elliptic curve is bi-rationally equivalent to such a Weierstrass equation.}
\beq
  \cC_p: \qquad  y^2 = x^3 + f\, x \, z^4 + g \, z^6 \ ,
  \label{WeierForm}
\eeq
where $x$, $y$, $z$ are projective coordinates 
in $\mathbb P^2_{2,3,1}$ and $f,g$ depend on the location $p$. Away from non-singular points on the base,
$f(u)$, $g(u)$ are holomorphic in the complex base coordinates $u$. When the elliptic curve 
becomes singular, the discriminant given by
\beq \label{eq:discriminant}
   \Delta = 4 f^3 + 27 g^2 
\eeq
vanishes. The vanishing of this function describes complex co-dimension one space in $B_3$ 
and determines the location of the space-time filling seven-branes on $B_3$.

Recall that F-theory on 
an elliptically fibered Calabi-Yau fourfold yields minimally supersymmetric 
theory in four space-time dimensions.\footnote{In fact, 
one could study the theory on the space $Y_4$ obtained 
by resolving the orbifold singularities of $Y^{\rm s}_4$ in a way compatible with the 
Calabi-Yau condition and the elliptic fibration.} 
After quotienting by the involution $\s$ and carrying out the resolution supersymmetry will be broken
by the geometry. 

The involutive symmetry $\sigma$ on the elliptic fibration is demanded to have a definite action on $B_3$, i.e.~$\sigma$ 
is compatible with the fibration and induces a well-defined action on the base that we also denote by $\sigma$ for simplicity. 
In a given local patch $U$ on $B_3$ described by the coordinates $(z_1, z_2, z_3)$ this action can be of different types 
with differing dimension of the fixed space $L_\sigma (U) \subset U$. For example, one has
\ba
(z_1, z_2, z_3) & \ra (\bar z_1, \bar z_2, \bar z_3)\,,  &\Rightarrow& \quad  \text{$L_\sigma(U)$ is a real 
three-dimensional subspace of $U$,}  \nn \\
(z_1, z_2, z_3) & \ra (\bar z_2, - \bar z_1, \bar z_3)\, ,  &\Rightarrow& \quad \text{$L_\sigma(U)$ is a real 
one-dimensional subspace of $U$,}  \nn \\
(z_1, z_2, z_3) & \ra \Big(\fr{\bar z_2}{\bar z_3}, - \fr{\bar z_1}{\bar z_3}, - \fr{1}{\bar z_3}\Big) \, ,&\Rightarrow& \quad \text{$L_\sigma(U)$ is empty and $\s$ is freely acting on $U$.}
\label{FixedPointsOnB}
\ea 
After taking the quotient the fixed space of $Y_4$ will represent orbifold singularity of $Z_8$ which must be resolved when moving to $\resZ_8$.  If $L_\sigma(B_3)$ is one dimensional then $\s$ is only an involution if $B_3$ already has an identification under the holomorphic orbifold action generated by $\s^2$. This will have a real two-dimensional fixed space that will be associated with an additional orbifold singularity over the base that must also be resolved after the quotient. 

The fixed space of $Y_4$, which we will call $L_\sigma(Y_4)$, can have components that are either $0$, $2$ or $4$ real dimensional 
or $\s$ can be freely acting. To investigate the action of $\sigma$ on $Y_4$ further we must analyze several cases which are distinguished by the location of the point $p$ on $B_3$:
\begin{itemize}
 \item[(1)]  \phantomsection \label{situ1} $p \nin L_\sigma(B_3)$: For each point $p$ on $B_3$ that is not a fixed point of $\sigma$ the corresponding
elliptic curve $\cC_p$ is mapped onto another elliptic curve $\cC_{\sigma(p)}$ over the 
image point $\sigma(p)$. However, since $\sigma$ is anti-holomorphic 
the orientations of $\cC_{\sigma(p)}$ and $\sigma(\cC_p)$ will differ. In this case $\s$ will be freely acting on all points of $Y_4$
that project to $p$ or $\sigma(p)$, see figure~\ref{Fig:exchange_torus}.

\begin{figure}[h]
\begin{center}
\includegraphics[width=8cm]{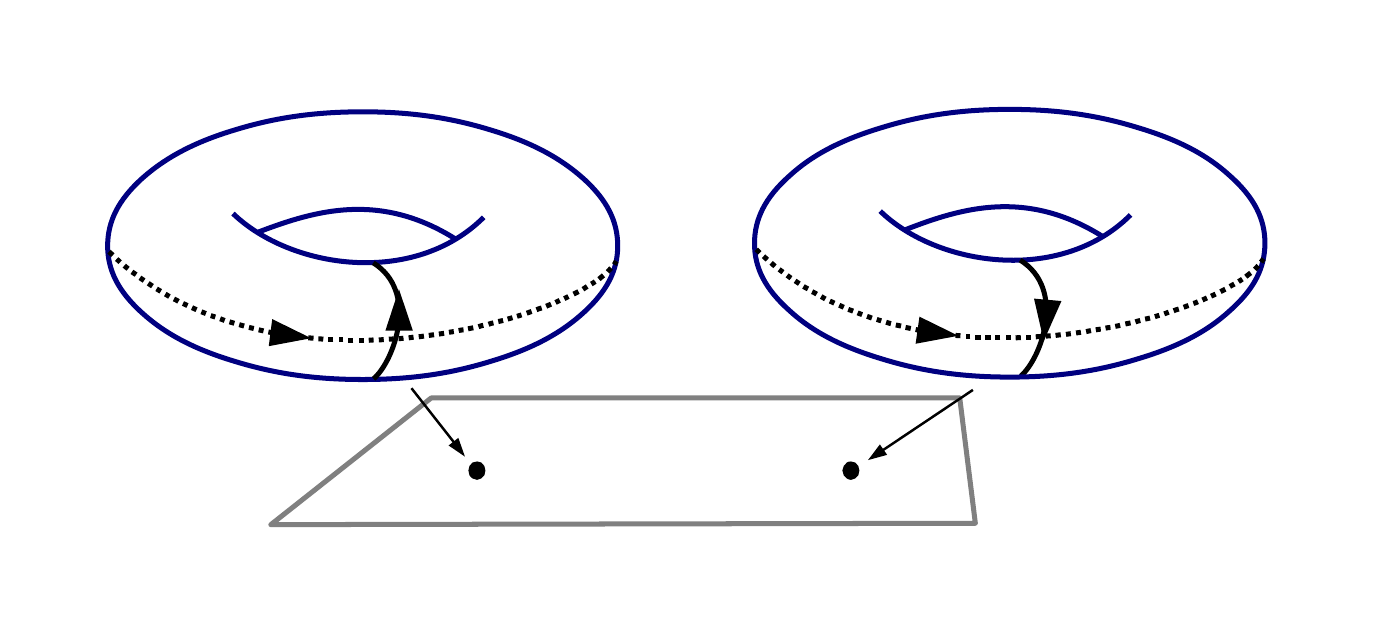}
\caption{Generic torus fibers exchanged by the anti-holomorphic involution.}
\label{Fig:exchange_torus}
\end{center}
\begin{picture}(0,0)
\put(140,95){$\cC_{p}$}
\put(335,95){$\cC_{\sigma(p)}$}
\put(212,78){$p$}
\put(241,78){$\sigma(p)$}
\end{picture}
\end{figure}
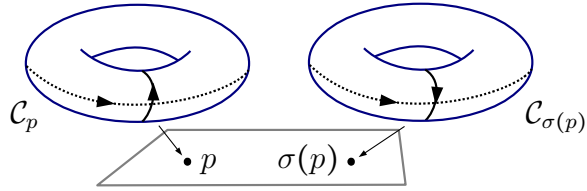

 \item[(2)] $p \in L_\sigma(B_3)$ and $ \Delta(p) \neq 0$: If a point $p$ on $B_3$ is a fixed point of $\sigma$
 the elliptic curve over this point will be mapped to itself. In particular, this implies that if $p$ is not on a seven-brane 
 that a smooth two-torus is mapped onto itself. Recall that the fixed point set of an anti-holomorphic involution 
on a smooth complex two-torus either consists of up to two real lines or is empty. 
\begin{itemize}
\item[(2.1)] \phantomsection \label{situ21} If the torus is fixed point free this implies that each point on $Y_4$ that projects to $p$
is actually not fixed by $\sigma$ and hence does not give rise to a singularity of $Z_8$. This means that $\s$ will be freely acting on
all points of $Y_4$ that project to $p$. If $L_\s(B_3)$ is one-dimensional then the additional singularities associated with the $\s^2$ identification can be resolved in the standard toric way. Interestingly, if $\s$ is fixed point free on the torus but not on the base then the quotient fiber at such $p$ is a Klein bottle, see figure~\ref{Fig:Kleinb}. 

\begin{figure}[h]
\begin{center}
\begin{minipage}[b]{0.42\textwidth}
\includegraphics[width=6.6cm]{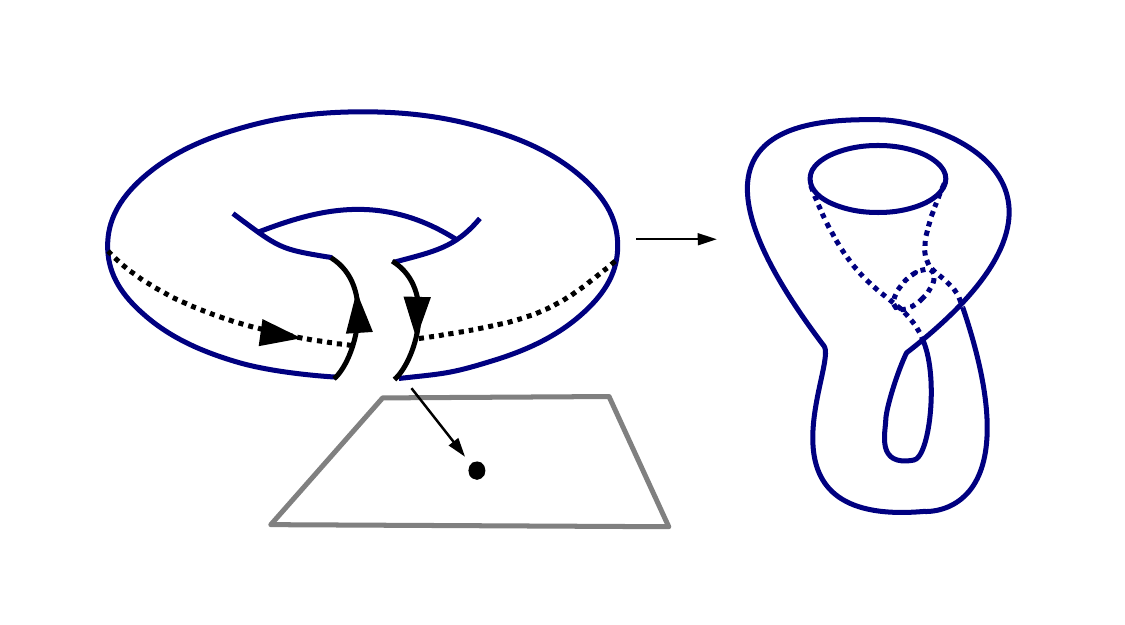}
\vspace*{-.3cm}
\caption{Fiber modded by anti-holomorphic involution
to Klein bottle fibers.}
\label{Fig:Kleinb}
\begin{picture}(0,0)
\put(5,110){$\cC_{p}$}
\put(168,110){$\cC_{p}/\sigma$}
\put(84,89){$p$}
\end{picture}
\end{minipage}
\hspace*{.8cm}
\begin{minipage}[b]{0.3\textwidth}
\includegraphics[width=4.5cm]{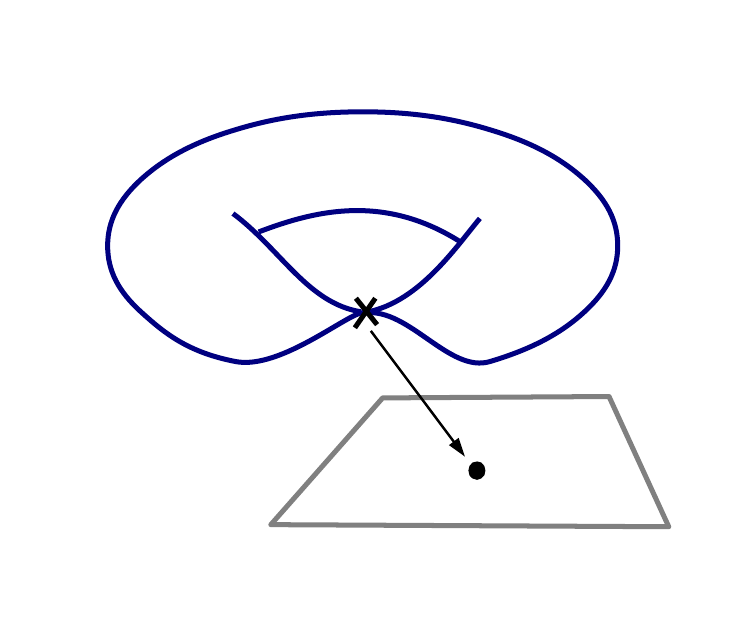}
\vspace*{-.5cm}
\caption{Nodal fiber at fixed point $p$. Involution fixes pinch-point.} \label{Fig:pinchedtorus}
\begin{picture}(0,0)
\put(5,110){$\cC_{p}$}
\put(86,85){$p$}
\end{picture}
\end{minipage}
\end{center}
\end{figure}

\item[(2.2)] \phantomsection \label{situ22} If the torus has a fixed line on it then the dimension of $L_\sigma(Y_4)$ may be up to one greater than the dimension of $L_\sigma(B_3)$, depending on the dimension of the subspace of $L_\sigma(B_3)$ over which the fixed space on the torus is a line. Since $L_\sigma(Y_4)$ must then have even dimensions greater than one, 
it must have dimension of either $2$ or $4$. The quotient of the elliptic curve by $\s$ then gives rise to a cylinder. \end{itemize}

 \item[(3)] \phantomsection \label{situ3} $p \in L_\sigma(B_3)$ and $ \Delta(p) = 0$: The most interesting case is if a point $p$ on $B_3$ is both a fixed point of 
 $\sigma$ and lies on a seven-brane. In this case $\cC_p$ is actually a singular curve. There are various possibilities for such singular curves 
 and a systematic study should investigate all possible anti-holomorphic involutions and their fixed points. Here, let us only consider 
 the simplest case where $\cC_p$ is a nodal curve ($I_1$ type), as schematically depicted in figure~\ref{Fig:pinchedtorus}. In this case there 
 can exist an involution $\sigma$ that has one fixed point exactly at the node of the elliptic curve. One can think of this nodal point  
 as arising by shrinking the real one-dimensional fixed point set of an anti-holomorphic involution on a smooth elliptic curve. In this case the dimension of $L_\sigma(Y_4)$ may be an even integer less than the dimension of $L_\sigma(B_3)$, so it can be either $0$ or $2$. 
\end{itemize}

From this we see that if the action of $\s$ on $Y_4$ is to be fixed point free then it can have only points for which situations \hyperref[situ1]{(1)} or \hyperref[situ21]{(2.1)} apply. Alternatively if we restrict the fixed space to consist only of isolated fixed points, which is imposed in \cite{Joyce:1999nk}, then we find that situation \hyperref[situ3]{(3)} must apply in which the torus is pinched at these points. In addition to this if we also wish to consider fixed points which are already holomorphic orbifold singularities of $Y_4$, as is also imposed in \cite{Joyce:1999nk}, then we find that $L_\sigma(B_3)$ must be one-dimensional. An example of a space which has singularities of this sort is shown in Appendix \ref{sec:CIExample}.

Let us now analyze the action of the anti-holomorphic involution
$\sigma$ on the elliptic fiber. 
To this end,
we consider the case in which the elliptic fibration is presented in Weierstrass form
\eqref{WeierForm} and we let the anti-holomorphic involution $\sigma$ act
anti-linearly on the projective coordinates of $\mathbb P^2_{2,3,1}$. Any $\sigma$ action of this type may then be brought into the form
\beq \label{FiberAction}
\sigma: \quad (x,y,z) \rightarrow (\bar x, \bar y, \bar z) 
\eeq
by an appropriate coordinate redefinition. 
Comparison between \eqref{WeierForm} and \eqref{FiberAction}
reveals that, in order for the anti-holomorphic involution to be well-defined
on the Calabi-Yau fourfold $Y_4$,
the sections $f$ and $g$ have to satisfy
\beq
f_{\sigma(p)} = \overline{f_p} \ , \qquad
g_{\sigma(p)} = \overline{g_p} \ ,
\eeq
for every $p$ on the base $B_3$. 
Recall that the  
modular parameter $\tau$ of the elliptic fiber
is given by
\beq
j(\tau) = \frac{4 \cdot (24 f)^3}{\Delta} \ ,
\eeq
where the discriminant was defined in \eqref{eq:discriminant}. We conclude that
for any point $p$ on the base $B_3$ 
\beq
j(\tau_{\sigma (p)}) = \overline{j(\tau_p)} = j(- \, \overline{\tau_p})\ .
\eeq
In the last step we have made use of the fact that 
the $j$-function admits a Laurent series in the variable $q = e^{2 \pi i \tau}$
with integer coefficients. In summary,  we can infer that
\beq \label{eq:tau_transf}
\tau_{\sigma (p)} = - \, \overline{\tau_p} \qquad
\text{up to $SL(2,\mathbb Z)$ transformations.}
\eeq
Note that this condition is perfectly compatible with a non-trivial
holomorphic dependence of the modular parameter on the base coordinates.
In particular, it can be satisfied for $\tau$ profiles with non-trivial
monodromies associated to the presence of seven-branes. 
Only in the special case in which $\tau$ is constant over the base, as in the 
weak coupling limit away from orientifold planes,
\eqref{eq:tau_transf} enforces a reality condition on $\tau$,
which has to be purely imaginary. We will comment on this further in 
section \ref{sec:weak+matter}.

\section{M-theory on Spin(7) Spaces and Calabi-Yau Quotients}

Having discussed the geometry of the Spin(7) holonomy manifolds that we wish to consider, we will now describe the effective theories which arise in the reduction of M-theory on these spaces. In subsection \ref{sec:MonSpin7} we will begin this analysis by considering the reduction on general Spin(7) manifolds. Then in subsection \ref{sec:MonSpin7fromCY} we will analyze how this may be related to the quotient of the effective theories that arise from compactification on Calabi-Yau fourfolds.  In subsection \ref{sec:ellipticSpin7} we will then restrict to the case where these Calabi-Yau manifolds are elliptically fibered and study the redefinitions that must be made in order to move into a frame that can be lifted to the 4d F-theory dual. 

\subsection{Effective Action of M-theory on Spin(7) Manifolds}
\label{sec:MonSpin7}

The compactification of M-theory on a Spin(7) manifold $\resZ_8$
yields a 3d effective theory with minimal $\cN=1$ supersymmetry. 
The action, to quadratic order in the fermions, for a \textit{general} 3d theory with $\cN=1$ supersymmetry 
can always be written in the form \cite{deWit:2003ja,deWit:2004yr}
\ba \label{eq:genearl3dN=1action}
S^{(3)}_{\cN=1} &=\int d^3 x\; e \,\bls 
\fr12 R - \fr14 \Theta_{IJ} \e^{\m \n \r} A^I_\m (\pa_\n A^J_\r + \fr13 {f_{K L}}^J A^K_\n A^L_\r ) -\fr12 g_{ \Lambda \Sigma} \cD_\m  \phi^\Lambda  \cD^\m \phi^\Sigma - V(\phi) \nn \\
& \quad
-\fr12 \bar \p_\m \g^{\m \n \r} D_\n \p_r 
-\fr12  g_{\S \L} \bar \c^\S \g^\m \cD_\m \c^\L   
 + \fr12 g_{\S \L } \bar \c^\S  \g^\m \g^\n \p_\m \cD_\n \f^\L \nn \\
&\quad 
- \fr1{2} F \bar \p_\m \g^{\m \n} \p_\n + \pa_\L F \bar \p_\m \g^\m \c^\L + \fr12 (g_{\S \L} F -2 D_\S \pa_\L F + 2 X_{\S}^{I} X_\L^J \Q_{IJ} )  \bar \c^\S \c^\L 
\brs \ ,
\ea
with covariant derivatives and scalar potential given by 
\ba \label{eq:gaugingandscalarpot}
     \cD_\mu \phi^\Lambda &= \partial_\mu \phi^\Lambda + \Theta_{IJ} X^{I \Lambda} A^I_\mu\ , &
      V(\phi) &= 2 g^{\Lambda \Sigma} \partial_\Lambda F \partial_\Sigma F - 4 F^2 \, . 
\ea 
Here $X^{I \Lambda}$ is the Killing vector of the target space symmetry 
that is gauged via \eqref{eq:gaugingandscalarpot}. The action \eqref{eq:genearl3dN=1action} contains the $\phi^\Lambda$-dependent metric 
$g_{\Lambda \Sigma}(\phi)$ that is non-degenerate and positive definite.
The coefficient $\Theta_{IJ}$ of the Chern-Simons term is symmetric in $I,J$, 
and constant which ensures the gauge invariance of the action. This represents the embedding 
tensor for the 3d gauged supergravity theory. 
The real function $F(\phi)$ depends on the scalars $\phi^\Lambda$ 
and is required to satisfy $\Theta_{IJ} X^{I\Lambda} \partial_\Lambda F = 0$ for gauge invariance.
 
For smooth Spin(7) geometries $\resZ_8$ the $\cN=1$ vacua where 
studied in  \cite{Papadopoulos:1995da,Becker:2000jc,Martelli:2003ki,Tsimpis:2005kj}.  
The 3d effective theory can be derived by reducing the action for 
11d supergravity \cite{Cremmer:1978km}, the bosonic part of which at lowest order in derivatives is given by 
\beq \label{11Daction}
S^{(11)} = \int \tfr12  R * 1 - \tfr14 G_4 \we * G_4 
           - \tfr1{12} C_3 \we G_4 \we G_4 \, , 
\eeq
as discussed in \cite{Becker:2000jc,Gukov:2001hf,Acharya:2002vs,Becker:2003wb}. 
In the full reduction one must also take into account the higher derivative terms along with the tadpole cancellation condition which for backgrounds without M2-branes becomes
\ba 
\frac{\chi(\resZ_8)}{24} = \frac12 \int_{\resZ_8} G_4 \wedge G_4 \ .
\label{eq:tadpole}
\ea
We will describe this 
reduction in the following and reconsider some aspects of the derivation 
presented in \cite{Becker:2003wb}. 
We stress that this reduction is actually a warped compactification, and 
we will neglect this back-reaction in the following leading order analysis.

We carry out the reduction by decomposing the metric and three-form of 11d supergravity as 
\ba
   ds^2  &= g_{\m \n} dx^\m dx^\n + g_{m n} dy^m dy^n \,, &
    C_3 &= A^I \wedge \omega_I\, ,
\ea
where $g_{mn}$ is the metric on $\resZ_8$ and $\omega_I$ form a basis for 
 $H^2(\resZ_8,\mathbb{R})$ with $I = 1, \ldots , b^2(\resZ_8)$. We will restrict to the 
case of $b^3(\resZ_8)=0$ for simplicity.
The 3d theory 
will then admit  U(1) gauge symmetries associated with the vectors $A^I$. 

In performing the Kaluza-Klein reduction one has to allow the metric 
of the internal geometry $\resZ_8$ to vary without leaving the 
class of Spin(7) geometries. To find the permitted deformations
one constructs the Lichnerowicz operator on $\resZ_8$ and shows that 
its zero modes are in one-to-one correspondence with the set of anti-self-dual 
four-forms $\xi_A, \, A = 1, \ldots, b^4_{\rm A}(\resZ_8)$, along with one additional zero mode that corresponds to a rescaling of the overall volume.
This implies that there will be $b^4_A(\resZ_8) + 1$ real scalar fields 
$\varphi^A$ and $\Vz$ parameterizing the deformations 
of the Spin(7) structure. The under a variation of the scalars $\Vz$ and $\varphi^A$ the Cayley calibration $\F$ and the metric 
are deformed as
\ba \label{4FormModuliDef}
\d \F &= K_{\Vz} \Phi\ \d \Vz + (K_A \F + \x_A)\ \d \vf^A  \ ,&
\d g_{m n} &= \fr1{4 \Vz} g_{m n} \d \Vz + \fr{7}{6 || \Phi ||^{2}} \, (\xi_{A})_{m p q r } \Phi_{n}{}^{p q r}  \ \d \vf^A
\ ,
\ea
where the factors in $\d g_{mn}$ are chosen in accord with \eqref{eq:useful}.
As a result of the anti-self-duality of $\xi_A$, 
the variation of the metric with respect to $\varphi^A$ is symmetric and trace-free \cite{Gibbons:1989er}. 
The real coefficients in \eqref{4FormModuliDef} given by $K_{{\Vz}}$ and $K_A$ are 
in general functions of $\Vz$ and $\varphi^A$ and depend on the normalization of 
$\Phi$.  

Upon performing the dimensional reduction, followed by a Weyl rescaling of the 
3d metric to move into the Einstein frame, the bosonic part of the effective action is given by
\beq \label{eq:3dN=1action_reduced}
   S^{(3)}_{\resZ_8} =\int \tfrac12 R *1 - \tfrac12 h_{IJ} F^I \wedge * F^J 
     - \tfr{1}{4} \Theta_{IJ} A^I \wedge F^J - \tfrac{1}{2} g_{\Vz \Vz} d \Vz \we * d \Vz -\tfrac12 g_{ A  B} d \varphi^{A} \wedge * d\varphi^{B} - V(\varphi) *1\ ,
\eeq
where
\ba
  g_{\Vz \Vz} = \tfr{9}{8} \Vz^{-2} \ , & &
  g_{AB} &=- \frac{7}{2}\frac{\int_{\resZ_8} \xi_{A} \wedge \xi_B}{\int_{\resZ_8} \Phi \wedge \Phi }\ ,& 
  h_{IJ} & =  \frac{1}{2\Vz} \int_{\resZ_8} \omega_{I} \wedge * \omega_J \, ,
\label{eq:Spin7ModMetric}
\ea
and the scalar potential $V(\varphi)$ is of the form \eqref{eq:gaugingandscalarpot}. This action is less general then \eqref{eq:genearl3dN=1action}. Firstly, 
we have only included Abelian vectors. More importantly, we did not dualize all dynamical vector degrees 
of freedom into scalar degrees of freedom as it is always possible in three dimensions. Therefore the kinetic terms of the vectors with 
$\varphi^{A}$-dependent metric $h_{IJ}$ still appears in \eqref{eq:3dN=1action_reduced}. Dualizing 
all vector degrees of freedom yields new scalars $\zeta_I$ with metric $h^{IJ}$, the inverse of $h_{IJ}$. 
The presence of a Chern-Simons term in \eqref{eq:3dN=1action_reduced}  implies that the 
$\zeta_I$ are in general gauged with covariant derivative 
\beq \label{eq:zeta_gauging}
   \cD \zeta_I =  d \zeta_I + \Theta_{IJ} A^J\ .
\eeq
Hence, the action \eqref{eq:3dN=1action_reduced} allows us to determine all couplings in \eqref{eq:genearl3dN=1action}: $\phi^\Lambda = (\Vz, \varphi^{ A},\zeta_I)$, 
$g_{\Lambda \Sigma}= (\fr{9}{8 \Vz^2}, g_{A B},h^{IJ})$, and $X^{I}_J =\delta_J^I,\, X^{I A}=0$. 

So far we have not discussed the scalar potential $V$ and the Chern-Simons coupling 
$\Theta_{IJ}$. In fact, in a compactification without fluxes both vanish identically. 
They are, however, induced if one allows for a non-trivial flux background of the field strength 
$dC_3$. Let us denote the background flux on $\resZ_8$ by $G_4$. A direct reduction of 
11d supergravity then implies that a flux-induced 
Chern-Simons term takes the form
\beq \label{eq:ThetaG4}
   \Theta_{IJ} = \int_{\resZ_8} G_4 \wedge \omega_I \wedge \omega_J\ .
\eeq
More involved is the derivation of the flux-induced scalar potential from a real function $F$. 
After dimensional reduction of the full action including the higher curvature term, one uses 
the tadpole cancellation condition \eqref{eq:tadpole} to show that the scalar potential takes the form 
\beq \label{eq:scalar_pot}
  V = \frac{1}{4\Vz^3} \Big( \int_{\resZ_8} G_4 \wedge * G_4 - \int_{\resZ_8} G_4 \wedge G_4 \Big) = - \frac{1}{2 \Vz^3} \int_{\resZ_8} G^{A}_4 \wedge  G^A_4 \ , 
\eeq
where $G^A_4$ is the anti-self-dual part of the background flux $G_4$.
To generally derive $F$ let us first note that it was argued in \cite{Becker:2003wb}
that $F$ should be proportional to $\int_{\resZ_8} G_4 \wedge \Phi$.
The factor in front of this flux integral can, however, be field-dependent.
In fact the correct form of $F$ is given by 
\ba \label{eq:general_F}
   F& =  \fr{\sqrt{7}}{4 \sqrt{2}  || \F|| \Vz^2 } \int_{\resZ_8} G_4 \wedge \Phi \, , 
\ea
The derivatives of $F$ then satisfy  
\ba \label{eq:derivativesF}
  \frac{\partial F}{\partial \varphi^A} &=  \fr{\sqrt{7}}{4 \sqrt{2} || \F|| \Vz^2 } \int_{\resZ_8} G_4 \wedge \xi_A  \ , &
  \frac{\partial F}{\partial \Vz} &=-\frac{3}{2 \Vz} F\ ,
\ea
which are independent of the precise form of $K_{\Vz}$ and $K_A$ in \eqref{4FormModuliDef} as these 
cancel when taking the derivative.\footnote{
One can also show that given a general Cayley calibration $\F$, which varies as \eqref{4FormModuliDef}, it is possible to define an alternatively normalized self-dual four-form $\hat \Phi$ which is also a singlet of Spin(7) and satisfies
\ba \label{eq:formofKA}
  \hat \Phi &=\fr{1}{  || \F|| \Vz^2 } \Phi \, , &
    \hat K_{\Vz} &=- \tfrac32 \Vz^{-1}\ , & 
    \hat K_A &= 0\ .
\ea
This corresponds to the normalization for $\F$ chosen in \eqref{eq:Phi_preliminary}.
}
Inserting \eqref{eq:derivativesF}, \eqref{eq:general_F} and the inverse metrics $g^{AB},g^{\Vz\Vz}$ 
obtained from \eqref{eq:Spin7ModMetric} into the general form of the $\cN=1$ scalar potential \eqref{eq:gaugingandscalarpot}
one readily shows match with \eqref{eq:scalar_pot}.

We conclude this section
by performing a rearrangement
of the Spin(7) moduli that will be useful in the
comparison to the Calabi-Yau reduction of section \ref{sec:MonSpin7fromCY}.
To begin with, we divide  the 
Spin(7) moduli $\varphi^A$ into two subsets,
$\varphi^A = ( \varphi^\cK, \varphi^{\til I_-})$. This  notation is chosen to make contact
to  section \ref{sec:MonSpin7fromCY}. 
Note that this partition of the Spin(7) moduli is supposed  to be such that
the associated anti-self-dual four-forms satisfy the
orthogonality condition 
\beq \label{eq:wedge-zero}
\int_{\hat Z_8} \xi_\cK \wedge \xi_{\tilde I_-} = 0 \ .
\eeq
Next we extend the range of the index $\til I_-$ by defining a new index $I_-$ that includes one additional entry and define $\phi^{I_-} = (\hat \phi , \hat \phi \varphi^{\tilde I_-})$. This definition is such that that the variation of $\F$ in  \eqref{4FormModuliDef} is now given by 
\ba
\d \F &= K_{\Vz} \Phi\ \d \Vz + (\mathbb K_{I_-} \F + \h_{I_-} ) \d \f^{I_-}  + (K_{\cK} \F + \x_{\cK} ) \d \vf^{\cK}  \ ,
\ea
where 
\ba
\mathbb K_{I_-} & = \bigg(- \frac{\varphi^{\tilde J_-}  K_{\tilde J_-}}{\hat \phi}  , 
\frac{K_{\tilde I_-}}{\hat \phi}
\bigg)\, ,  & 
\eta_{I_-} &= 
\bigg(  - \frac{\varphi^{\tilde J_-} \xi_{\tilde J_-}}{\hat \phi} , \frac{\xi_{\tilde I_-}}{\hat \phi} \bigg) \ .
\ea
These definitions then imply the constraints
\ba
\f^{I_-} \mathbb K_{I_-} & = 0 \ , & 
\phi^{I_-} \, \eta_{I_-} = 0 \ ,
\ea 
which means that the action  \eqref{eq:3dN=1action_reduced} develops a new local symmetry under under which 
\ba
\f^{I_-} &\ra \l \f^{I_-}\, ,  &
\F &\ra \l \F\, . 
\label{extrasymmetry} 
\ea 
As anticipated above, this constrained formulation
will be helpful in section \ref{sec:MonSpin7fromCY}. It might
also be useful, however,
 in finding generalizations
of the F-theory construction to Spin(7) manifolds that are not obtained
as Calabi-Yau quotients.

\subsection{Effective Action of M-theory on Spin(7) Manifolds from
Calabi-Yau Quotients} \label{sec:MonSpin7fromCY}

In the following we would like to introduce Spin(7) geometries whose 
effective theories can be up-lifted to four dimensions via 
the M-theory to F-theory limit. It is an outstanding 
question to characterize such geometries generally. In order to 
approach this problem we therefore restrict our analysis to Spin(7) geometries 
arsing from elliptically fibered Calabi-Yau fourfolds as 
introduced in Section \ref{sec:spin7fromCY}. Our aim is to first show, that 
the 3d $\cN=2$ theories arising 
in Calabi-Yau fourfold compactifications of M-theory 
are truncated to $\cN=1$ when performing the anti-holomorphic 
quotient $Y_4/\sigma$, with an involution $\sigma$ as in \eqref{eq:OmegaJtransform}.
We note that the following steps bear many similarities 
to the construction of 4d Type IIA Calabi-Yau 
orientifold actions \cite{Grimm:2004ua}. However, here we are truncating 
3d $\cN=2$ supersymmetry to $\cN=1$ supersymmetry.\footnote{A 
systematic study of spontaneous $\cN=2$ to $\cN=1$ breaking in three dimensions can 
be found in \cite{Hohm:2004rc}.} Truncations of $\cN=2$ Chern-Simons theories 
to $\cN=1$ induced by an anti-holomorphic involution have been also considered in \cite{Forcella:2009jj}.

Let us first recall the general form of a 3d $\cN=2$ action. 
The bosonic part of this can always be brought to the form 
\beq \label{eq:genearl3dN=2action}
   S^{(3)}_{\cN=2} =\int \tfrac12 R *1 - \tfrac14 \Theta_{IJ} A^I \wedge (d A^J + \tfrac23 {f_{K L}}^J A^K \wedge A^L) 
   - g_{ \cA \bar \cB} \cD M^\cA \wedge * \cD \bar M^\cB - \tilde V *1\ ,
\eeq
where $g_{ \cA \bar \cB} = \partial_{\cA} \partial_{\bar \cB} K$ is a K\"ahler metric and
$\tilde V(M,\bar M)$ is the scalar potential. 
This scalar potential is generally of the form 
\beq
   \til V = e^{K} \big(K^{\cA \bar \cB} D_\cA W \overline{D_\cB W} - 4 |W|^2 \big) + \big( K^{\cA \bar \cB} \partial_\cA \cT \overline{\partial_\cB \cT} - \cT^2 \big)  \ , 
\eeq
where $W(M)$ is a holomorphic superpotential and $\cT$ is a real potential. One may also note that 
in the $\cN=2$ case the presence of a non-vanishing $\cT$ is linked to the gaugings $\cD M^\cA$.

The 3d $\cN=2$ effective action 
for a Calabi-Yau fourfold compactification of 11d supergravity 
was derived in \cite{Haack:1999zv,Haack:2001jz}. For the case $b^3(Y_4)=0$ it takes a particularly simple 
form. The reduction yields $h^{3,1}(Y_4)$ complex structure moduli $z^\cK$, 
which are complex fields and parametrize the changes of the $(4,0)$-form $\Omega(z)$.
In addition there are $h^{1,1}(Y_4)$ real K\"ahler structure deformations $v^I$ 
arising in the expansion of the K\"ahler form $J=v^I \omega_I $. 
The expansion of the M-theory three-form $C_3=A^I\wedge  \omega_I$ yields $h^{1,1}(Y_4)$ 
3d vectors $A^I$. The vectors $A^I$ together with $v^I$ form the bosonic 
components of 3d $\cN=2$ vector multiplets. After dualizing all dynamical vector 
degrees of freedom into scalars $\zeta_I$, the kinetic terms of the 
3d $\cN=2$ supergravity theory are encoded 
by a K\"ahler potential 
\ba \label{eq:def-KMtheory}
   K(z,T) & = - \log \int_{Y_4} \Omega \wedge \bar \Omega  - 3 \log \Vy \, , 
\ea
which is evaluated as a function of the $h^{3,1}(Y_4)+h^{1,1}(Y_4)$ complex coordinates $z^\cK$ and 
\beq \label{eq:def-TI}
    T_I = \tfrac{1}{3!} \int_{Y_4} \omega_I \wedge J^3 + i \zeta_I \ .  
\eeq
In the presence of background fluxes $G_4$ a non-trivial Chern-Simons term 
with $\Theta_{I J }$ exactly as in \eqref{eq:ThetaG4} is induced. As above in \eqref{eq:zeta_gauging} this 
also implies the presence of gaugings $\cD T_{I} = dT_{I} + i \Theta_{I J } A^J$. 
Furthermore, a scalar potential arises from the functions
\ba \label{eq:def-cTW}
   \cT &= \frac{1}{4 \Vy^2 }\int_{Y_4} G_4 \wedge J^2\ , &
    W &= \int_{Y_4} G_4 \wedge \Omega  \ ,
\ea
where $\cT$ is in accord with the gauged shift symmetries.

In order to implement the $\cN=1$ truncation we first note 
that the relevant forms have to transform under 
$\sigma^*$ as 
\ba \label{eq:trans_JCOmegaC3}
   \sigma^*J &= -J \ ,& 
   \sigma^*(C \Omega) &=  \overline{C \Omega}\ ,& 
   \sigma^* C_3 &= C_3\ ,
\ea
where the first two conditions already appeared in \eqref{eq:OmegaJtransform} 
when inserting the definition
\beq
   C = e^{-i\theta} e^{K/2}\ ,
\eeq
with $K$ as defined in \eqref{eq:def-KMtheory}. 
To perform the reduction one thus has to 
split the cohomology of $Y_4$ into parity-even 
and parity-odd eigenspaces as
\beq
  H^{n}(Y_4,\mathbb{R}) = H^{n}_+(Y_4,\mathbb{R}) \oplus H^n_-(Y_4,\mathbb{R})\ . 
\eeq
The surviving vectors in the expansion of $C_3$ only arise from elements of $H^2_+(Y_4)$,
while the surviving K\"ahler structure scalars arise from elements of $H^2_-(Y_4)$.
Thus, one has
\ba
   C_3 &= A^{I_+} \wedge \omega_{I_+} \ , & I_+ &= 1, \ldots , h^{1,1}_+(Y_4)\ , & 
       J & =  v^{I_-} \omega_{I_-}\ ,        & I_- &= 1, \ldots , h^{1,1}_-(Y_4)  \ .  &  
\ea
Applying this to the dual complex scalars $T_I$ introduced in \eqref{eq:def-TI}
one finds the split 
\ba
  T_I &= (T_{I_+} , T_{I_-})  = (-i \text{Im} T_{I_+}, \text{Re} T_{I_-}) \ , &
   \text{Im} T_{I_-} &= \text{Re} T_{I_+} = 0 \ . 
\ea
In other words, out of the $h^{1,1}(Y_4)$ complex coordinates $T_I$ only $h^{1,1}(Y_4)$ 
\textit{real} coordinates survive in the quotient theory. 
Similarly, the $h^{3,1}(Y_4)$ complex fields  $z^\cK$ encoding complex structure deformations 
are reduced to $h^{3,1}(Y_4)$ real complex structure deformations $\vf^\cK$.
This can be inferred by considering all complex structure deformations of $\Omega$ preserving 
the condition \eqref{eq:trans_JCOmegaC3}. One can chose local coordinates such that $ \varphi^\cK = \R\, z^\cK$.
In summary, the involution truncates the $\cN=2$ K\"ahler manifold spanned by $T_I$ and  $z^\cK$
to a real Lagrangian submanifold $\cL_\sigma$ parametrized by $\zeta_{I_+}$, $\text{Re} T_{I_-}$ and $\varphi^\cK$.

To compare these degrees of freedom which survive the quotient with 
those described in the Spin(7) reduction of subsection \ref{sec:MonSpin7} 
it is necessary to redefine the fields. The vectors $A^{I_+}$ and 
the volume $\Vy$ are simply identified with the vectors $A^I$ and 
the volume $\Vy$ in \eqref{eq:3dN=1action_reduced}, while the $b^4_A(Z_8)$ scalar 
fields $\varphi^{A}$ in \eqref{eq:3dN=1action_reduced} parametrize the independent degrees of freedom of the constrained fields
\ba 
    \f^{\hat A} &= (\vf^\cK,\vn^{I_-} )\ , &
    & \text{where} & 
    \hat A &= 1, \ldots, 1 + b^4_A(Z_8) \,, &
    \vn^{I_-} &= \Vy^{-\tfr14} v^{I_-}\ .
\label{eq:def-varphiCY}
\ea
They satisfy the constraint 
\beq \label{eq:normalizev}
    \Vn  \equiv  \tfr{1}{4!}  \cK_{I_- J_- K_- L_-}   \vn^{I_-} \vn^{ J_-} \vn^{K_- } \vn^{L_-} = 1 \, , 
\eeq
as a result of the definition \eqref{eq:def-varphiCY}. This condition can be viewed as a gauge fixing of the additional symmetry introduced in \eqref{extrasymmetry}.
In terms of these fields the bosonic part of the effective theory describing the projected Calabi-Yau reduction is given by
\ba
   S^{(3)}_{Y_4 / \s} & =\int \tfrac12 R *1 - \tfrac12 h_{I_+ J_+ } F^{I_+} \wedge * F^{J_+}
     - \tfrac14 \Theta_{I_+ J_+ } A^{I_+} \wedge d A^{J_+}  
- \tfr12 g_{\Vy \Vy} d \Vy \we * d \Vy \nn \\
& \quad -\tfrac12 \til g_{ I_- J_- } d \vn ^{I_-} \wedge * d \vn  ^{J_-}   - \tfr12 \til g_{ \cK \cI } d \vf^\cK \wedge * d \vf^\cI - V *1\ ,
\label{QuotientedY4action}
\ea 
where the scalar metrics may be written as
\ba
g_{\Vy \Vy} & = \tfr{9}{8} \Vy^{-2} \, , & 
h_{I_+ J_+} & =  \frac{1}{2\Vy} \int_{Y_4} \omega_{I_+} \wedge * \omega_{J_+}\,, \nn \\
\til g_{ I_- J_- } & = - 4 \Vy^3  \int_{Y_4}  \eta_{I_-} \we \eta_{J_-} ,  & 
\til g_{ \cK \cL }  & = - 4 \Vy^3 \int_{Y_4} \x_{\cK} \we \x_{\cL} \,, & 
\label{truncatedCYModuliMetric}
\ea 
and where 
\ba
\eta_{I_-} & = \tfr{1}{4} \Vy^{-\fr32}  P_{I_-}{}^{J_-} \o_{J_-} \we \Jn \ ,& 
& P_{I_- }{}^{J_-}  = \d_{I_-}{}^{J_-} - \fr{1}{4!}  \cK_{I_- K_- L_- M_-} \vn ^{K_- }\vn ^{L_-} \vn ^{M_-}  \vn ^{ J_-} \, ,  \nn \\ 
 \x_\cK & =   \Re (C \c_{\cK} )   \ ,& 
& \cK_{I_- J_- K_- L_-} = \int_{Y_4} \o_{I_-} \we \o_{ J_- } \we \o_{K_- } \we \o_{L_-}  \,. 
\label{xiFromCY}   
\ea
We have used the definition $ \Jn = \vn ^{I_-} \o_{I_-}$. Note that the constraint \eqref{eq:normalizev} is responsible for the projection 
matrices $P_{I_-}{}^{J_-}$ that appear in the definition of the scalar metric. The Chern-Simons terms in \eqref{QuotientedY4action}
are induced by $G_4$ fluxes as in \eqref{eq:ThetaG4} and read
\beq \label{Theta++}
   \Theta_{I_+ J_+} = \frac{1}{2}\int_{Y_4} \omega_{I_+} \wedge \omega_{J_+} \wedge G_4\ .
\eeq

By considering the potential of the truncated theory and matching this with \eqref{eq:gaugingandscalarpot} we see that 
\beq  \label{eq:F_splitted_on_quotient}
   F = e^{K/2} \text{Re} W + \tfrac{1}{2} \cT = \int_{Y_4} G_4 \wedge \big(\text{Re} (C \Omega) + \tfrac18 \Vy^{-2} J \wedge J\big) \, .
\eeq
By comparing this with \eqref{eq:general_F} we may then read off $\F =  \big( \text{Re} (C \Omega) + \tfrac18 \Vy^{-2} J \wedge J\big)$\ up to a choice of normalization.
This is the expression for $ \Phi$ that we already quoted in \eqref{eq:Phi_preliminary}. 
In the remainder of this subsection we discuss the structure of 
the resulting Spin(7) field space in more detail. 

To investigate the metric on the Spin(7) field space 
we need to determine its variations with respect to the coordinates 
introduced in \eqref{eq:def-varphiCY}. This again requires the constraint \eqref{eq:normalizev}
to be consistently implemented.
One way to achieve this is to first express $ \Phi$ in terms of $\Vy$
and $\Vn $ before taking derivatives and later impose \eqref{eq:normalizev}.
Concretely, one has 
\beq
  \F = \frac{1}{\Vy^{3/2}} \left( \frac{\text{Re} (e^{-i\theta} \Omega)}{\big(\int_{Y_4} \Omega \wedge \bar \Omega \big)^{1/2}}+ 
           \frac{1}{8} \frac{\Jn \wedge \Jn}{\Vn ^{1/2}}\right)\ .
\eeq
Then taking the variations of this with respect to $\Vy$, $\vn ^{I_-}$, and $\vf^\cK$ we find
\ba
\d \F|_{\Vn =1} = - \tfrac32 \Vy^{-1} \, \Phi\ \delta \Vy +\eta_{I_-} \  \d \vn ^{I_-} 
+ \x_\cK \ \d \vf^{\cK}  \, , 
\label{VarPhiFromCY}
\ea 
and in addition find that the normalization of $\F$ is such that 
\ba
\int_{\resZ_8} \F \we \F = \tfr{7}{16} \Vy^{-3} \ .
\label{CYPhiNorm}
\ea
Then by comparing the variation \eqref{VarPhiFromCY} with \eqref{4FormModuliDef} we may identify 
the forms $\xi_\cK$ and $\eta_{I_-}$ with the Spin(7) forms $\xi_A$. More precisely,
note that the constraint \eqref{eq:normalizev} implies $\phi^{I_-} \, \eta_{I_-} = 0$.
We thus identify the coordinates $\phi^{I_-}$ and forms $\eta_{I_-}$ with the 
quantities constructed after   \eqref{eq:wedge-zero}.
Moreover, we find that the projected $Y_4$ moduli metric \eqref{truncatedCYModuliMetric} matches the Spin(7) moduli metric \eqref{eq:Spin7ModMetric}. As expected from the general Spin(7) analysis, $\eta_{I_-}$  and $\x_\cK$ also form a basis for the complete set of anti-self-dual four-forms of $Y_4$ which are invariant under $\s \, $.\footnote{
In fact the basis formed by $\eta_{I_-}$  and $\x_\cK$ is complete but also degenerate as a result of the  projection matrix $P_{I_-}{}^{J_-}$ which appears in the definition of $\eta_{I_-}$.
}

\subsection{Effective Action of M-theory on Spin(7) Quotients of Elliptically Fibered Calabi-Yau Fourfolds} 
\label{sec:ellipticSpin7}

In order to derive the 4d effective action of F-theory 
on a Spin(7) holonomy manifold, we must now restrict our M-theory reduction of 
section \ref{sec:MonSpin7fromCY} to be based on elliptically fibered Calabi-Yau fourfolds.
In doing this we will denote the base of the elliptically fibered Calabi-Yau $Y_4$ by $B_3$. 
Recall that for an elliptic fibration we find in cohomology 
\beq
    12 c_1(B_3) = [ \Delta ]_{B_3}\ ,
\eeq
where $[\Delta]_{B_3}$ is the Poincar\'e-dual two-form 
to the discriminant locus \eqref{eq:discriminant} in the base $B_3$. 
We note that both $c_1(B_3)$ and $[\Delta]_{B_3}$ have to 
transform with a negative sign under the anti-holomorphic and isometric involution $\sigma$. This 
requirement also ensures that $\Delta$ has a finite volume, i.e.~$\int_{\Delta} J \wedge J$ does not vanish.

The two-form 
associated to the zero section of the elliptic fibration is denoted by $\omega_0$. 
In this work we will be only dealing with Calabi-Yau fourfold geometries with 
holomorphic zero sections. 
Note that $\omega_0$ must transform with a negative sign under $\sigma^*$. In 
fact, as we discussed in section \ref{sec:spin7fromCY} the homology 
class of the torus fiber is negative under $\sigma$, since $\sigma$ reverses the 
orientation of the two-torus. This property can also be seen by noting that the base intersects the fiber exactly once. As we will discuss later, this allows us 
to perform the uplift by sending the coefficient $\f^0$ in the expansion of $J$ 
to zero. 

As the involution $\s$ also descends to the base, the cohomology of $B_3$ may be decomposed under the action of $\s$ as $H^{p}(B_3) = H^{p}_+(B_3) \oplus H_-^{p}(B_3) $. This means that one can write 
\beq \label{eq:base_split}
   (\omega_\alpha)=(\omega_{\alpha_+},\omega_{\alpha_-})\ ,  \qquad \alpha_\pm = 1,\ldots, h^{1,1}_\pm(B_3)\ ,
\eeq
where $\omega_{\alpha_{\pm}}$ are obtained by pulling back elements of $H^{2}_{\pm} (B_3)$ to $H^{2}_{\pm} (Y_4)$.

We will also allow for resolved singularities of 
the elliptic fibration of $Y_4$ that correspond to simple non-Abelian gauge groups $G$ in the dual F-theory compactification on $Y_4$. 
The location of these non-Abelian singularities defines a divisor $S$ in $B_3$. In the simple analysis that follows we will assume that there is only one stack of seven-branes on $B_3$ that describe a non-Abelian gauge group and so $S$ has only one connected component. This significant simplification by no means represents the most general setup which we will not address here. As a result the actions that follow will not represent the most general possibilities.

The Poincar\'e dual two form $[S]_{B_3}$ lifted to $Y_4$, admits 
the expansion $b^{\alpha_-}_S \omega_{\alpha-}$ defining 
constant coefficients $b^{\alpha_-}_S$.
As noted above, $[\Delta]_{B_3}$ and hence $[S]_{B_3}$ have negative parity under $\sigma$ so only the $\omega_{\alpha-}$
appear in the expansion.  The non-Abelian singularities are resolved 
by introducing new two-forms $\omega_i,\ i =1, \ldots , \text{rank}(G)$.
Assuming the absence of Abelian gauge factors one has $\text{rank}(G) = h^{1,1}(Y_4) - h^{1,1}(B_3)-1$.
Let us note that all $\text{rank}(G)$ forms $\omega_i$ are in fact negative under 
$\sigma^*$. To infer this we stress that each exceptional divisor is 
a $\mathbb{P}^1$-fibration over the seven-brane locus in the base $B_3$. 
Within $B_3$ the seven-brane divisor $S$ and its volume form 
are positive under $\sigma$ by Poincar\'e duality.\footnote{Recall 
that formally $\sigma (B_3) = - B_3$, since $\sigma$ reverses the orientation of $B_3$.} 
Since the anti-holomorphic $\sigma$ reverses the sign of the volume form of the 
$\mathbb{P}^1$-fiber, we conclude that the exceptional divisors and their Poincar\'e dual 
two-forms $\omega_i$ are negative under $\s$. 
In summary, we find that the two-forms representing $H^{2}(Y_4)$ are split 
according to 
\beq  \label{omega+-split}
      (\omega_{I_+}) = (\omega_{\alpha_+})\ ,\qquad   (\omega_{I_-}) = (\omega_0,\omega_{\alpha_-},\omega_{i})\ .
\eeq
This implies that the truncated spectrum of the 3d $\cN=1$ theory 
is given by $h^{1,1}_+(B_3)$ vectors $A^{\alpha_+}$, and $h^{1,1}(Y_4)-h^{1,1}_+(B_3) + h^{3,1}(Y_4)$ 
scalars $v^{I_-} = (v^0 , v^{\alpha_-},v^i)$ and $\varphi^\cK$.

One can now systematically study all intersection numbers that are not forbidden by 
the $\sigma$-parity. Since the volume form on $Y_4$ is positive under $\sigma^*$
the vanishing intersection numbers $\cK_{IJKL} = \int_{Y_4} \omega_I \wedge \omega_J \wedge \omega_K \wedge \omega_L$ are 
\beq
  \cK_{I_+ J_+ K_+ L_-} =0 \ , \qquad   \cK_{I_+ J_- K_- L_-} = 0 \ .
\eeq
Combined with the intersection structure on elliptic fibrations 
one thus finds that   for the  potential $\hat K = K|_{\cL_\sigma}$ the relevant non-vanishing intersections 
are
\ba \label{eq:relevant_intersect}
   \cK_{0 \alpha_- \beta_- \gamma_-} &\equiv \kappa_{\alpha_- \beta_- \gamma_-} \ , & \cK_{0 \alpha_- \beta_+ \gamma_+} &\equiv  \kappa_{\alpha_- \beta_+ \gamma_+}\ ,& \\
 \cK_{ij \alpha_- \beta_-} &= - C_{ij} b^{\gamma_-}_S \kappa_{ \gamma_- \alpha_- \beta_-} \ , & 
  \cK_{ij \alpha_+ \beta_+} &= - C_{ij} b^{\gamma_-}_S \kappa_{ \gamma_- \alpha_+ \beta_+} \ , & \nn 
\ea
where $\kappa_{\alpha_- \beta_- \gamma_-}$ and $\kappa_{\alpha_- \beta_+ \gamma_+}$ are the 
triple intersections on $B_3$. The matrix $C_{ij}$ is the Cartan matrix of the non-Abelian gauge group $G$. 
Let us stress that there are numerous other intersection numbers that are in general non-zero on $Y_4/\sigma$.
In particular, intersection numbers involving $(\omega_0)^n,\, n>0$ will play a crucial role 
when matching the F-theory and M-theory reduction at the one-loop level
\cite{Grimm:2011fx,Cvetic:2012xn,Cvetic:2013uta}.\footnote{They can 
be reduced by repeatedly using $(\omega^0)^2 = - c_1(B_3) \wedge \omega_0$.} 
Crucially, this requires a redefinition of the coordinates 
\ba
\hat \phi^{\alpha_-} = \phi^{\alpha_-} + \frac12 K^{\alpha_-} \phi^0,
\label{hatAlphaRedef}
\ea 
where $-K^{\alpha_-}$ are the coefficients of $c_1(B_3)$ in the basis $\omega_{\alpha_-}$ \cite{Grimm:2011sk}. 

The splitting of the $v^{I_-}$ coordinates then induces a splitting of the constrained Spin(7) 
moduli $\f_{I_-}$ defined in \eqref{eq:def-varphiCY}. 
After performing the redefinition \eqref{hatAlphaRedef} we may then move into a set of redefined 
coordinates that are appropriate for performing the F-Theory lift.
Firstly, $\phi^0$ is mapped the length of the interval and we set
\ba
   \fr{1}{r^2}  & = \f^0\cV^{-\fr34}\ ,
\ea
where $r$ is the circumference of the circle in $S^1/\mathbb{Z}_2$. Hence, $\f^0$ 
captures degrees of freedom of the 4d metric.  
The $\hat \phi^{\alpha_-}$ become 4d scalars, while 
the $\phi^i$ are the scalar part of 4d vectors with index 
along the interval $\phi^i_{\rm b} = A^i_3$.
It is convenient to set
\ba \label{eq:fieldredefinitions}
\f_{\rm b}^{\a_-} & = ({\f^0})^\fr13 \wh \f^{\a_-} -\tfr12 ({\f^0})^{-\fr{2}{3}} b^\a C_{i j} \f^i \f^j \ , & 
\cV_{\rm b}  & = ({\f^0})^{\fr1{2}} \cV^{\fr98} \ , & 
\phi^i_{\rm b} &=   ({\f^0})^{- 1}  \f^i \ .
\ea
These redefinitions can 
be motivated by the fact that, when taking the F-theory limit with large $r$, 
the constraint \eqref{eq:normalizev} only depends on $\f_{\rm b}^{\a_-}$, while $r$ and $\phi^i_{\rm b}$
are unconstrained. In addition, following \cite{Grimm:2010ks} the vectors $A^{\alpha_+}$ will become 4d scalars with a real shift symmetry. 
We will consider the lift more explicitly in section \ref{sec:4daction}.

Let us finally also consider the flux-induced Chern-Simons couplings $\Theta_{I_+ J_+}$ and potential $F$, given in \eqref{Theta++} and \eqref{eq:F_splitted_on_quotient}. From the 
split \eqref{omega+-split} we infer that the Chern-Simons coupling $\Theta_{\alpha_+ \beta_+}$ only involves vectors that become 4d
scalars and therefore, by the considerations of \cite{Grimm:2011sk}, have to be absent
\beq
    \Theta_{\alpha_+ \beta_+} = 0\ . 
\eeq 
The real potential $F$ can be expressed in terms of $\Theta_{I_- I_J}$
as 
\beq
     F = \int_{Y_4} G_4 \wedge \R (C\Omega) + \tfrac18 \cV^{-1} \Theta_{I_- J_-} \phi^{I_-} \phi^{J_-}\ .
\eeq
Again using \eqref{omega+-split} and following \cite{Grimm:2011sk} one has to additionally impose 
\beq \label{vanishing_Theta}
    \Theta_{0 0} = 0\ , \quad \Theta_{0 \alpha_-} = 0 \ ,  \quad \Theta_{0 i} = 0\ ,  \quad \Theta_{\alpha_- \beta_-} = 0\ ,\quad \Theta_{i \beta_-} = 0\ .
\eeq
This choice of fluxes allows that a 4d theory might exist, no fluxes are 
included in reduction from four to three dimensions, and the gauge-group $G$ is un-broken in four dimensions.\footnote{These 
conditions will be modified in the presence of U(1) gauge factors \cite{Grimm:2011fx,Cvetic:2012xn,Cvetic:2013uta}.}
The resulting potential $F$ will contain a term that is classical on the F-theory side 
and a one-loop contribution as we will discuss at the end of the next section.

\section{F-theory on Spin(7) Manifolds}

In the previous section we studied M-theory on Spin(7) manifolds and 
later focused on examples constructed as quotients of  elliptically 
fibered Calabi-Yau fourfolds
by an anti-holomorphic involution. 
As a next step we discuss in subsection \ref{sec:interval_reduction}
the dual interval reduction of a 4d theory. Concretely, 
we will identify the boundary conditions on various  
4d fields on an interval that have to be imposed 
in order to make a duality of the form \eqref{new_dual} possible. 
Aspects of the non-supersymmetric 4d effective theories 
are discussed in subsection \ref{sec:4daction}. We particularly focus 
on the couplings of the uncharged scalar fields that are real 
both in three and four dimensions and satisfy Neumann
boundary conditions at the ends of the interval.

\subsection{Dimensional Reduction of the 4d Theory on an Interval} \label{sec:interval_reduction}

One of the crucial ingredients of the new kind of M-theory/F-theory 
duality claimed in \eqref{new_dual}
is the use of an interval in the dimensional reduction from
four to three dimensions on the F-theory side of the duality.
In this subsection we discuss some general features of dimensional reduction
on an interval and consider candidate 4d parent actions.

Due to the presence of an interval $I=S^1/\bZ_2$ in \eqref{new_dual} the up-lift  
of a 3d theory on $\cM_3$ to a 4d theory on $\cM_4 = \cM_3 \times I$ is further complicated, 
since boundary conditions have to be given for each field.
These have to be appropriately specified in order 
that the duality suggested in \eqref{new_dual} holds. 
In the following we will discuss vectors, fermions, and 
scalars in turn. 

Let us first consider a  4d Abelian vector
$ A_m$.
Since its components satisfy a second-order equation of motion
we can choose Dirichlet or Neumann conditions. This choice,
however, has to be such that each component of the field strength
$F_{mn}$ has a definite parity under the $\mathbb Z_2$ action.
In particular, inspection of the 
the mixed component
\beq
 F_{\mu 3} = \partial_\mu A_3 - \partial_3 A_\mu
\eeq
reveals that if $ A_\mu$ satisfies Dirichlet boundary conditions
$ A_3$ has to satisfy Neumann boundary conditions, and vice versa. This gives the two choices 
\bea \label{eq:vector_boundaries}
 \text{(A)}&\qquad& D:\quad  \left. A_\mu \, \right|_{\partial \cM_4} = 0 
\qquad
\text{and}
\qquad N:\quad
\left. \partial_3 \, A_3 \,  \right|_{\partial \cM_4} = 0 \ ,\\
\text{(B)}&\qquad& D:\quad  \left. A_3 \, \right|_{\partial \cM_4} = 0 
\qquad
\text{and}
\qquad N:\quad
\left. \partial_3 \,  A_\mu \,  \right|_{\partial \cM_4} = 0\ , \nn
\eea
that may be made without over constraining the equation of motion. When carrying out the interval reduction the Dirichlet boundary conditions will remove the would-be zero mode of the corresponding 4d field. So fields with Dirichlet boundary conditions will not be seen in the 3d effective theory. 
This implies that reduction of $A_m$ can yield either one massless scalar or one massless
vector in the 3d effective action, but not both. 
This fact can be extended to non-Abelian gauge fields 
for a 4d gauge group $G$. To do this let us denote the generators 
of the algebra of $G$ by $(T_i, T_{\cI})$, with $T_i$ labeling the  
Cartan generators. Then for each vector
$A^i_m, A^\cI_m$ one can choose different boundary conditions. 

To conform with the theory arising in the Spin(7) reduction 
it turns out that one needs to chose option (A) in \eqref{eq:vector_boundaries}
for the Cartan vectors to keep 3d scalars $\phi^i_{\rm b} = A^i_3$ 
and option (B) for the non-Cartan vectors in order to keep 3d vectors $A^\cI_\mu$.\footnote{These boundary conditions imply that the gauge coupling constant should be effectively assigned odd parity under the $\bZ_2$ action.}
In this case one notes that the non-Cartan 3d vectors $A^\cI_\mu$
acquire a mass term for which the mass is determined by the vacuum expectation value of the 3d massless scalars $\phi^i_{\rm b}$.
This mass term arises in the effective theory from the reduction of the gauge kinetic term. This analysis is consistent with the fact that 
the 3d theory arising in the reduction described in section \ref{sec:ellipticSpin7} is a Wilsonian effective action with no non-Cartan vectors and only the scalars $\phi^i,\, i=1,\ldots, \text{rank}(G)$.
Let us stress, however, that we are still able to extract the classical couplings 
using the Spin(7) reduction by uplifting the couplings of the 
scalars $\phi^i_{\rm b}$. The Lorentz transformations and gauge transformations 
of the 4d vector mix all components of $ A_m^i, A^\cI_m$ and thus allow to recover the 
couplings of the 4d vectors from the couplings of $\phi^i_{\rm b}$, for a large interval 
on which these symmetries are restored.

Let us next consider a 4d fermion given 
by a Majorana spinor $\chi$.
Since its equations of motion are first-order,
we can only impose a Dirichlet boundary condition of the form
\beq
\left. \tfrac 12 (1 \pm \gamma^3) \chi \, \right|_{\partial \cM_4} = 0
\eeq
without over constraining the dynamics. The sign is related to the intrinsic parity
of the spinor under the $\mathbb Z_2$ action on the interval. 
For both choices, reduction of $\chi$ furnishes a massless Majorana
spinor in the 3d effective action. 
This implies that when focusing on zero modes, the degrees of freedom 
of the fermions are halved. However, there is no ambiguity 
when uplifting a fermion from three to four dimensions. 4d
Lorentz invariance implies that the 3d dynamics of the spinor 
encodes its 4d couplings. A similar argument applies to the 
gravitino.

The comparison 
can, however, be more involved if the 4d fermion is \textit{charged} 
under the gauge group $G$. 
In an interval reduction the Coulomb branch 
scalars can give dimensionally reduced fermions a mass proportional to $\phi^i_{\rm b}$ 
if the coupling to $\phi^i_{\rm b}$ is non-vanishing. This implies 
that these fermions are not part of the low-energy effective theory 
and have to be integrated out.
As with the vectors we find that the Cartan fermions remain dynamical in the 3d low-energy 
effective theory. These then comprise the 3d, $\cN=1$ 
supersymmetric partners of $\f^i_{\rm b}$ moduli.

Finally, we turn to the reduction of a 4d scalar field $\phi$
with standard two-derivative action yielding a 
second-order equation of motion. As a result, we can impose Dirichlet
or Neumann boundary conditions
\beq
\left. \phi \, \right|_{\partial \cM_4} = 0 
\qquad
\text{or}
\qquad
\left. \partial_3 \,  \phi \,  \right|_{\partial \cM_4} = 0
\eeq
without over constraining the equation of motion.  As a result 
the degree of freedom of a 4d scalar might be entirely lost (for Dirichlet b.c.) 
or kept (for Neumann b.c.) when considering only the zero mode
in the 3d effective theory. This is in contrast to the vectors and fermions discussed 
above. In other words, one can add an arbitrary number 
of Dirichlet scalars to a candidate 4d action without 
changing the 3d effective theory on a small interval.

These features of interval reductions lead us to 
first specify a minimal 4d Lorentz invariant 
ansatz for the 4d action containing only those couplings 
that can be uniquely fixed by comparison with 
the 3d $\cN=1$ zero mode action. 
This non-supersymmetric theory is given to 
quadratic order in the fermions by
\ba \label{eq:4dactionansatz}
  S^{(4)}_{\rm Min}& = \int d^4 x \; e\, \Big[
  - \tfr12 R - \tfr12\, \cG_{\cA \cB}\, \partial_m \vf^\cA \partial^m \vf^\cB 
  - \tfr14 f \, \Tr( F_{m n} F^{m n} ) - V^{(4)} \nn \\
  & \quad
  - \tfr12 \bar \p_m \g^{m n r} D_n \p_r 
  - \tfr12 \cG_{\cA \cB} \bar \c^\cA \g^m D_m \c^\cB 
  - \tfr12 f\, \Tr( \bar \l \g^m D_m \l ) 
  + \tfr14  f \, \bar \p_m \g^{r s} \g^{m} \Tr (\l F_{rs}) 
  \nn \\
  & \quad 
  + \tfr1{2 \sq2 } \cG_{\cA \cB} \bar \p_m \g^n \g^m  \c^\cA D_n \vf^\cB  
  + \tfr12  A^1 \bar \p_m \g^{m n}  \p_n 
 + \tfr{1}{\sq{2}} A^2_\cA \bar \p_m \g^m  \c^\cA 
  \nn \\ 
  & \quad
- \tfr12  A^3_{\cA\cB} \bar \c^\cA  \c^\cB 
  +  \tfr{1}{4 \sq{2}} A^4_\cA\,  \Tr( F_{m n}  \bar \l ) \g^{m n} \c^\cA 
- \tfr12  \cG^{\cA \cB} A^4_\cA\, A^2_\cB \Tr(\bar \l \l) 
 \Big]\, ,
\ea
where the covariant derivatives of the Majorana fermions are given by 
\ba
D_m \p_n & = \pa_m \p_n + \tfr14 \o_{m r s} \g^{r s} \p_{n} \, , & 
D_m \l & = \pa_m \l  + \tfr14 \o_{m r s} \g^{r s} \l   + [A_m,\l]\,, \nn 
\ea
\vspace{-0.8cm}
\ba
D_m \c^\cA & = \pa_m \c^\cA + \tfr14 \o_{m r s} \g^{r s}  \c^\cA  + D_m \f^\cB \G_{\cB \cC}{}^{ \cA} \c^\cC \, .
\ea
In this action $G_{\cA \cB}$ is a real metric for the scalar target space and $V^{(4)}$, $f$ are real functions of the scalars $\varphi^\cA$. In addition to this $A^1$, $A_\cA^2$, $A^3_{\cA \cB}$ and $A^4_\cA$ are further functions of $\varphi^\cA$ that will later be determined by comparing the reduction of this action with the 3d result. As this action is not supersymmetric we could in principle have made a much more general proposal for the couplings that appear. However, it will turn out that \eqref{eq:4dactionansatz} is sufficiently general to allow for a matching with the 3d theory to be performed. For convenience we note here that performing this calculation one finds that the potential is given in terms of a real function $\cF$ by  
\ba \label{gen4Dpotential}
V^{(4)} = 2 G^{\cA \cB} \partial_\cA \cF \partial_\cB \cF - 3 \cF^2\ ,
\ea 
and that the $A$ functions are given in terms of $\cF$ and $f$ by
\ba \label{Afunctions}
A^1 & = \cF \, , &
A^2_{\cA} & = \pa_\cA \cF \,, &
A^3_{\cA \cB} & = D_\cA \pa_\cB \cF - \tfr12 \cG_{\cA \cB} \cF \,.  &
A^4_{\cA} & = \pa_\cA f \, . 
\ea

The action $S^{(4)}_{\rm Min}$ given in \eqref{eq:4dactionansatz} should be used with caution.
Indeed its interpretation as a Wilsonian effective action 
is questionable, since it was obtained as 4d Lorentz convariantization of the 
3d effective theory that applies in the small interval limit. It was constructed 
as the minimal completion of the 3d action consistent with 4d Lorentz invariance. 
All scalars satisfying Dirichlet boundary conditions have only massive excitations and are dropped from 
the action \eqref{eq:4dactionansatz}.
These fields are strictly speaking not moduli even in the absence of fluxes. We will comment further on 
these Dirichlet scalars above, but will not discuss their impact in detail. They might, however, restore 4d, $\cN=1$ supersymmetry in 
the 4d bulk if the size of the interval is taken to infinity.\footnote{We are grateful to Eran Palti and Ralph Blumenhagen 
for useful discussions on this point.} 

A possible 4d Wilsonian effective action $S^{(4)}_{\rm W}$ completing $S^{(4)}_{\rm Min}$ on a large interval could be given by a 
$\cN=1$ Lagrangian $\cL^{(4)}_{\cN=1}$  for F-theory on the original Calabi-Yau space $Y_4$ 
supplemented by the boundary conditions or a boundary action $\cL^{(3)}$.
Hence, it takes the form 
\beq \label{4dN=1boundary}
    S^{(4)}_{\rm W} = \int_{\cM_4} \cL^{(4)}_{\cN=1} + \int_{\partial \cM_4} \cL^{(3)}\ .
\eeq
The restauration of the Calabi-Yau moduli space from the 
moduli space of the Spin(7) manifold in the large interval limit 
would be very non-trivial. In a follow-up paper we will explore this 
treatment further \cite{toappear}.

Let us stress that the action \eqref{eq:4dactionansatz}  neglects the couplings of charged matter that 
will be present in a general F-theory compactification. Furthermore, we have 
not displayed the terms of higher order in the fermions. These can be 
added by making an Ansatz for these couplings 
and reducing them to three dimensions with the boundary conditions 
described above. The coefficients are then determined by comparing the
zero mode result to a general 3d, $\cN=1$ theory in which the higher fermionic couplings are known in terms of the 3d $\cN=1$ characteristic 
functions determined by the reduction of the terms in \eqref{eq:4dactionansatz}.

As stressed above the minimal action \eqref{eq:4dactionansatz} could 
be modified by adding an arbitrary number of scalars 
satisfing Dirichlet boundary conditions without 
modifying the classical 3d low-energy effective action for the zero modes. 
The question of determining the true 4d Wilsonian action can thus be not 
resolved from a purely supergravity perspective.
A similar problem occurs for the ambiguities encountered in 
the up-lift of 3d $\cN=2$ scalars on a circle in the standard M-theory F-theory duality. 
In such an up-lift a 3d scalar can either be part of a 4d $\cN=1$ 
vector or chiral multiplet. The decisive information in determining 
this ambiguity arises form the M-theory to F-theory limit and 
the geometry of the Calabi-Yau fourfold. We will discuss this point further in 
what follows.

\subsection{Effective Action of F-theory on Spin(7) Manifolds} \label{sec:4daction}

Having described the 3d effective theory obtained
for the quotient torus fibered Spin(7) geometry in subsections \ref{sec:MonSpin7fromCY} and \ref{sec:ellipticSpin7} and 
the details on the interval reduction in subsection \ref{sec:interval_reduction} 
we are now in the position to perform the reduction and read off the
couplings of the 4d theory
\eqref{eq:4dactionansatz}. 
Clearly, proposing that the coupling functions take the same form in 
the 4d theory is a speculative part of the analysis. It 
amounts on the one hand 
to sending the size of the interval $I$ to infinity, and on the 
other hand shrinking the fiber volume. 
This means that 
one has to be performing the M-theory to F-theory limit. 
In supersymmetric F-theory compactifications it has 
become clear over the last years \cite{Grimm:2010ks,Grimm:2011fx,Bonetti:2011mw} that many 
couplings in the 3d theory obtained 
from M-theory appear to also have an F-theory interpretation. 
Motivated by these advances we perform a similar oxidation
for the Spin(7) compactification. However, it should be stressed 
that we will only talk about zero modes in the following and 
many of the subtleties are, in fact, hidden in the treatment of massive
modes. 

The first step is to implement the F-theory limit explicitly. 
Note that not all couplings arising in the M-theory reduction 
are classical from the F-theory perspective on a small compact space. 
Various couplings can be induced at loop level when integrating out massive Coulomb 
branch and Kaluza-Klein modes. To extract the classical terms only, 
one can assert scalings to the various fields as suggested in \cite{Grimm:2010ks}.
The correct scalings are \cite{Bonetti:2011mw}
\beq
   v^0 \ \rightarrow\ \epsilon v^0\ , \qquad v^{\alpha_-} \ \rightarrow \ \epsilon^{-1/2} v^{\alpha_-}\ , \qquad v^{i} \ \rightarrow \ \epsilon^{1/4} v^i \ , \qquad 
   r\ \rightarrow \ \epsilon^{-3/4} r\ .  
\eeq
They ensure precisely that the couplings with intersection numbers \eqref{eq:relevant_intersect}, i.e.~$\cK_{0\alpha_- \beta_-,\gamma_-}, \cK_{0\alpha_- \beta_+,\gamma_+}$ 
and $\cK_{ij \alpha_- \beta_-},\cK_{ij\alpha_+ \beta_+}$ are surviving  the $\epsilon \rightarrow 0$ limit. 
Translated into the coordinates $\phi^{I_-}$ one thus finds 
\beq \label{eq:eps_scaling}
   \phi^0   \ \rightarrow\ \epsilon^{9/8} \phi^0\ , \qquad \phi^{\alpha_-}\ \rightarrow\ \epsilon^{-3/8} \phi^{\alpha_-} \ ,\qquad \phi^{i}\ \rightarrow\ \epsilon^{3/8} \phi^{i}\ , \qquad 
   \cV \ \rightarrow \ \epsilon^{-1/2} \cV\ .
\eeq
Combining these scalings with the coordinate redefinitions \eqref{eq:fieldredefinitions}
one extracts the leading terms of all fields.
We first introduce the $\phi^{\alpha_-}_{\rm b}$ defined as 
the leading term in \eqref{eq:fieldredefinitions}. In the limit the normalization 
constraint \eqref{eq:normalizev} translates to the condition 
\beq
  N_{\rm b} \equiv \tfrac{1}{3!} \kappa_{\alpha_- \beta_- \gamma_-} \phi_{\rm b}^{\alpha_-} \phi_{\rm b}^{\beta_-} \phi_{\rm b}^{\gamma_-} =1 \ .
\eeq
This implies that only $h^{1,1}_-(B_3)-1$ coordinates $\phi^{\alpha_-}_{\rm b}$ are independent. The missing degree of freedom 
is encoded by the base volume $\cV_{\rm b}$ arising as leading term in the definition \eqref{eq:fieldredefinitions}.
After the $\epsilon \rightarrow 0$ limit the resulting 
3d action can be matched with a the reduction of  a 4d theory reduced on an 
interval of length $r$ with boundary conditions introduced in subsection \ref{sec:interval_reduction}. 
This allows us to read off the data of the 4d theory from the 3d action. 

We first note that all couplings containing 3d vectors or fermions are formally lifted 
from 3d to 4d in a Lorentz compatible way. For example, the kinetic terms in \eqref{eq:genearl3dN=1action} for the 3d fermions $\chi^{\alpha_-}$, which are in the 
same 3d, $\cN=1$ multiplets as the scalars $\phi^{\alpha_-}_{\rm b}$, are given by
\beq
   \tfrac{1}{2} \tilde g_{\alpha_- \beta_-} \bar \chi^{\alpha_-} \slashed{\cD} \chi^{\beta_-} \, . 
\eeq
These are lifted by completing the $\chi^{\alpha_-}$ into 4d fermions and matching $\tilde g_{\alpha_- \beta_-}$ with the reduction of the equivalent 4d terms after performing the reduction and Weyl rescaling as well as implementing 
the $\epsilon \rightarrow 0 $ limit with \eqref{eq:eps_scaling}. In this way we can read off 
\beq
\cG_{\a_- \b_-} = (\tilde g_{\alpha_- \beta_-}) _{\epsilon = 0} = 4  \cV^3_{\rm b} \int_{B_3} \xi^{\rm b}_{\alpha_-}  \wedge * \xi^{\rm b}_{\beta_-}\ ,
\eeq
where the four-forms $\xi^{\rm b}_{\alpha_-}$ are given by 
\ba
   \xi^{\rm b}_{\alpha_-} &= \tfrac{1}{4} \cV_{\rm b}^{-\fr43} {P_{\alpha_-}}^{\gamma_-} \omega_{\gamma_-} \wedge \omega_{\beta_-} \phi^{\beta_-}_{\rm b}\ , & 
   P_{\alpha_- }{}^{\beta_-} & = \d_{\alpha_-}{}^{\beta_-} - \tfr{1}{3!  }  \kappa_{\alpha_- \gamma_- \delta_-} \phi_{\rm b}^{\gamma_- }\phi_{\rm b}^{\delta_-}   \phi_{\rm b}^{\beta_-} \, .
\ea
The other components of the 4d scalar metric $G_{\cA \cB}$ appearing in \eqref{eq:4dactionansatz} may then be deduced in a similar way by expanding $\vf^\cA = ( \cV_b, \, \f^{\a_-}, \vf^\cK, \z_{\a_+} )$  and making the comparison with \eqref{eq:genearl3dN=1action} and \eqref{QuotientedY4action}. This gives  $\cG_{\cV_b \cV_b }  = \tfr4{6}\cV_{\rm b}^{-2}$ and 
\ba \label{eq:4dmetric}
\cG_{ \cK \cL } & = (\tilde g_{\cK \cL}) _{\epsilon = 0} = 4  \cV^3_{\rm b} \int_{B_3} \xi^{\rm b}_{\cK}  \wedge * \xi^{\rm b}_{\cL}  \, , &
 \cG^{\a_+ \b_+ } & = ( h_{\a_+ \b_+ } )^{-1}_{\epsilon = 0} =  (\frac{1}{2 \cV_b} \int_{B_3} \omega_{\a_+} \wedge * \omega_{\b_+})^{-1} \, , 
\ea

Next we can consider the comparison of the kinetic terms for the scalars $\f^i$ with the reduction of 
the 4d vector kinetic terms. In this way we find that the coupling function $f$ is given by 
\ba \label{eq:gauge_coupling_function}
f C_{i j} =  (r^2 g_{i j}) _{\epsilon = 0} =  \cV^{2/3}_{\rm b} C_{i j} b^{\alpha_-}_S \kappa_{\alpha_- \beta_- \gamma_-} \phi_{\rm b}^{\beta_-} \phi_{\rm b}^{\gamma_-} \ . 
\ea
Similarly the reduction of the potential for the 4d theory may be compared with the general 3d, $\cN =1$ result \eqref{eq:gaugingandscalarpot} from which we find \eqref{gen4Dpotential} where 
the function $\cF$ is related to the function $F$, which determines the potential of the quotiented Calabi-Yau reduction, by
\ba \label{eq:fluxF}
 \cF = (r F) _{\epsilon = 0} =  \Big(e^{K^{\rm F}/2} \int_{Y_4} \text{Re}(\Omega) \wedge G_4 \Big)_{\cL_\sigma} \ .
\ea
where $K^{\rm F} = - 2 \log \cV_{\rm b} - \log \int_{Y_4} \Omega \wedge \bar \Omega$. 
Finally we note that by comparing the fermionic couplings in the reduction of \eqref{eq:4dactionansatz} with \eqref{eq:genearl3dN=1action} we find \eqref{Afunctions}.  

We stress that in contrast to a supersymmetric effective theory the 
couplings of the bosons are less restricted and holomorphicity 
does neither protect the generating potential \eqref{eq:fluxF} nor the 
gauge coupling \eqref{eq:gauge_coupling_function}.
It would be desirable to check if 3d, $\cN=1$ supersymmetry 
helps to nevertheless ensures additional control over the 
corrections to these couplings on a finite size interval.

In the preceding analysis we did not include charged matter. 
Clearly, in a general F-theory compactification with 
fluxes chiral matter will be part of the 4d massless spectrum. 
This matter can become massive when dimensionally 
reduced on an interval if the scalars $\phi^i_{\rm b}$ get 
a vacuum expectation value. This implies that these have to 
be integrated out in the 3d low-energy effective theory.
In contrast to the 3d, $\cN=2$ theories arising 
in Calabi-Yau fourfold compactifications \cite{Grimm:2011fx,Cvetic:2012xn,Cvetic:2013uta} there is 
no one-loop contribution of chiral matter to 3d Chern-Simons 
terms in our 3d, $\cN=1$ setup. However, part of the 
3d potential $F$ will admit a one-loop term 
\beq
      F \supset F^{\rm class} + F^{\rm 1-loop} \ .
\eeq
This classical term will lift to the 4d superpotential \eqref{eq:fluxF}
in our simple configurations with only one unbroken non-Abelian 
gauge group. The one-loop term can be obtained by considering 
the general Spin(7) potential $\cF$ with \eqref{eq:F_splitted_on_quotient}, imposing 
that up-lift conditions \eqref{vanishing_Theta}, and keeping the term 
that vanish in the limit $\epsilon \rightarrow 0$. This leads to the identification 
\beq
   F^{\rm 1-loop}    \overset{?}{=}  \tfrac{1}{8} \cV^{-2} \int_{Z_4} J \wedge J \wedge G_4 =  \tfrac{1}{8} \cV^{-1} \Theta_{ij} \phi^i \phi^j\ .
\eeq
It would be very interesting to check this match for an explicit example by computing both the 
general one-loop contribution in field theory and the flux intersection $\Theta_{ij}$ of the 
form \eqref{eq:ThetaG4}.  

Let us close with a brief comment on the Kaluza-Klein modes 
in the interval reduction. In the M-theory to F-theory duality on elliptically 
fibered Calabi-Yau manifolds the Kaluza-Klein modes map 
to M2-branes that wrap also the elliptic fiber. This implies that these 
states are charged under the Kaluza-Klein vector. In quotient 
torus fibered Spin(7) manifolds it is therefore crucial 
to investigate M2-brane states wrapping the fiber. Since, in the 
generic case of figure \ref{Fig:exchange_torus}, the torus is mapped to 
an orientation reversed image, such M2-branes states appear in pairs. 
It remains to be checked which effects these have on the 
supersymmetry of the 3d and 4d effective theories. 
It is an important task to investigate these massive states 
to gain deeper insights into the correct choices of boundary 
conditions on the interval and the Kaluza-Klein 
compactification from 4d to 3d.

\section{Comments on Weak Coupling and Charged Matter} \label{sec:weak+matter}

This section is devoted to the discussion of some aspects
of the weak coupling limit for F-theory on Spin(7)
manifolds. In particular, we focus on the case in which the Spin(7)
manifold is a quotient torus fibration as described in section \ref{Geoms}.
We propose a Type IIB realization of the setup and we 
briefly comment on the charged matter spectrum in this 
string theory picture.

\subsection{Weak Coupling Interpretation}\label{sec:weak}

In what follows  we describe a proposal for the 
Type IIB realization of the weak coupling limit of F-theory
on Spin(7) manifolds constructed as anti-holomorphic quotients
of Calabi-Yau fourfolds.  

Before the anti-holomorphic involution is implemented we have
F-theory on the fourfold $Y_4$ with base $B_3$.
In the weak coupling limit \cite{Sen:1997gv} this becomes
a Type IIB orientifold on a Calabi-Yau threefold $Y_3$
acted upon by a holomorphic involution $\sigma_{\rm hol}$
in such a way that $B_3 = Y_3 /\sigma_{\rm hol}$.
After the implementation of the 
anti-holomorphic involution $\sigma$ on $Y_4$ on the M-theory
side it is natural to expect a further quotient of the Type IIB 
setup under the associated involution $\sigma$ acting on the base
$B_3$.
In summary, we propose that the weak coupling 
picture of the setup under examination is furnished by Type
IIB on the Calabi-Yau threefold $Y_3$ quotiented
both by a holomorphic involution 
and by an anti-holomorphic involution.

To make this proposal more precise we need to determine the 
intrinsic parities of Type IIB fields under the action of the 
anti-holomorphic involution. 
As in the standard discussion of the M-theory/F-theory
duality \cite{Denef:2008wq} it is convenient to start with the simple case of 
M-theory on a product manifold $\cM_9  \times T^2$.
The eleven-dimensional metric
takes the form 
\beq
ds_{11}^2 = \frac{v}{\tau_2} \left[ (dx + \tau_1 dy)^2 + \tau_2^2 dy^2 \right] +  ds_9^2  \ ,
\eeq
where $x,y$ are coordinates one the torus, 
which has modular parameter $\tau = \tau_1 + i \tau_2$.
M-theory is reduced on the circle parametrized by
$x$ to get Type IIA. The $y$-circle is the T-duality circle.
In this factorized case we consider an anti-holomorphic involution
$\sigma$ that acts separately on $\mathcal M_9$ and $T^2$.
More precisely, $\sigma$ acts on a complex three-dimensional
submanifold $B_3$ of $\mathcal M_9$ as described in section \ref{sec:spin7fromCY}.
Equation \eqref{eq:tau_transf} applied to the case of a constant
fibration shows that $\tau$ has to be purely imaginary in
order to have compatibility with the anti-holomorphic involution.
As a result, the anti-holomorphic action $z \rightarrow \bar z$,
where $z = x + \tau y$, is equivalent to $x \rightarrow x$,
$y \rightarrow  -y$.

The above observation is useful in establishing
the intrinsic parities of Type IIB fields under the action of 
$\sigma$ by means of the following heuristic argument.
To begin with, we  associate negative intrinsic parity to the $y$ coordinate and 
positive intrinsic parity to the $x$ coordinate.
Next, we observe that all M-theory fields have positive parity
under $\sigma$. Finally, we use the standard chain of dualities to
identify the M-theoretical origin of each Type IIB field and deduce its
parity.\footnote{Our application of Buscher's rules  
is purely schematic and is intended only as a tool to 
read off the $\sigma$-parities of Type IIB fields.} 
This step is summarized in Table \ref{table:parities}
together with our findings for the intrinsic $\sigma$-parities.

\begin{table}[h]
\centering
\begin{tabular}{ c c c c  }
\rule[.5cm]{0 cm }{0 cm}  IIB & 
\rule[.5cm]{0 cm }{0 cm}  IIA & 
\rule[.5cm]{0 cm }{0 cm} M & 
\rule[.5cm]{0 cm }{0 cm} $\sigma$-parity \rule[.5cm]{0 cm }{0 cm}  \\[.1cm] \hline\hline
\rule[.5cm]{0 cm }{0 cm}  $\phi$ & 
\rule[.5cm]{0 cm }{0 cm} $\phi$ & 
\rule[.5cm]{0 cm }{0 cm} $g_{xx}$ & 
\rule[.5cm]{0 cm }{0 cm} $+$\\[.1cm] \hline
\rule[.5cm]{0 cm }{0 cm}  $g_{\mu\nu}$ & 
\rule[.5cm]{0 cm }{0 cm} $g_{\mu\nu}$ & 
\rule[.5cm]{0 cm }{0 cm} $g_{\mu\nu}$ & 
\rule[.5cm]{0 cm }{0 cm} \multirow{3}*{\vspace{-.4cm}$+$}\\[.1cm]
\rule[.5cm]{0 cm }{0 cm}  $g_{\mu y}$ &
\rule[.5cm]{0 cm }{0 cm}  $(B_2)_{\mu y}$ & 
\rule[.5cm]{0 cm }{0 cm} $(C_3)_{\mu y x}$  \\[.1cm]
\rule[.5cm]{0 cm }{0 cm}  $g_{yy}$ & 
\rule[.5cm]{0 cm }{0 cm}  $g_{yy}$ & 
\rule[.5cm]{0 cm }{0 cm}  $g_{yy}$ \\[.1cm] \hline
\rule[.5cm]{0 cm }{0 cm}  $(B_2)_{\mu\nu}$ & 
\rule[.5cm]{0 cm }{0 cm}  $(B_2)_{\mu\nu}$ & 
\rule[.5cm]{0 cm }{0 cm}  $(C_3)_{\mu\nu x}$ & 
\rule[.5cm]{0 cm }{0 cm}  \multirow{2}*{\vspace{-.2cm}$-$}\\[.1cm]
\rule[.5cm]{0 cm }{0 cm}  $(B_2)_{\mu y}$ & 
\rule[.5cm]{0 cm }{0 cm}  $g_{\mu y}$ & 
\rule[.5cm]{0 cm }{0 cm}  $g_{\mu y}$ \\[.1cm] \hline
\rule[.5cm]{0 cm }{0 cm} $C_0$ & 
\rule[.5cm]{0 cm }{0 cm} $(C_1)_y$ & 
\rule[.5cm]{0 cm }{0 cm}  $g_{xy}$ & 
\rule[.5cm]{0 cm }{0 cm}  $-$\\[.1cm] \hline
\rule[.5cm]{0 cm }{0 cm}$(C_2)_{\mu\nu}$ & 
\rule[.5cm]{0 cm }{0 cm}  $(C_3)_{\mu\nu y}$ & 
\rule[.5cm]{0 cm }{0 cm}  $(C_3)_{\mu\nu y}$ & 
\rule[.5cm]{0 cm }{0 cm}  \multirow{2}*{\vspace{-.2cm}$-$}\\[.1cm]
\rule[.5cm]{0 cm }{0 cm}  $(C_2)_{\mu y}$ & 
\rule[.5cm]{0 cm }{0 cm}  $(C_1)_\mu$ & 
\rule[.5cm]{0 cm }{0 cm}  $g_{\mu x}$ \\[.1cm] \hline
\rule[.5cm]{0 cm }{0 cm}$(C_4)_{\mu\nu\rho y}$ & 
\rule[.5cm]{0 cm }{0 cm}  $(C_3)_{\mu\nu\rho}$ & 
\rule[.5cm]{0 cm }{0 cm} $(C_3)_{\mu\nu\rho}$ & 
\rule[.5cm]{0 cm }{0 cm} $-$
\end{tabular}
\caption{Schematic summary of type IIB fields with Type IIA duals, M-theory origin,
and intrinsic $\sigma$-parity.
Indices $\mu,\nu,\rho$ refer to the nine-dimensional manifold $\cM_9$, $x$ refers to the
direction of the M-theory
circle, $y$ refers to the direction of the circle along which T-duality is performed.}
\label{table:parities}
\end{table}

In summary, we conjecture that the Type IIB weak coupling picture 
of the Spin(7) compactifications we are studying is obtained
by taking the quotient under the symmetry group generated by the 
transformations
\beq
\cO_{\rm hol} = (-)^{F_L} \, \Omega_p \, \sigma_{\rm hol} \ , 
\qquad \qquad 
\cO = (-)^{F_L} \, P_3 \, \hat \sigma \ .
\eeq
The  expression for $\cO_{\rm hol}$ is the familiar orientifold action, with 
left-moving space-time fermion number $F_L$ and
world-sheet parity $\Omega_p$.
The  expression for $\cO$ deserves some comments.
Firstly, the inclusion of the factor $(-)^{F_L}$
is motivated by the intrinsic $\sigma$-parities of Table \ref{table:parities}.
Secondly, we have decomposed the action of $\sigma$ into 
two involutions $\hat \sigma$ and $P_3$.
On the one hand, the involution $\hat \sigma$ is the anti-holomorphic 
involution on $Y_3$ determined by the action of
$\sigma$ on $B_3$. 
On the other hand, the involution $P_3$ is the reflection of one spatial
 direction in $\mathbb R^{1,3}$,
\beq
P_3 : (x^0, x^1, x^2, x^3) \rightarrow (x^0, x^1, x^2, - x^3) \ .
\eeq
Without the factor of $P_3$ we would not have a  symmetry
of Type IIB. Indeed, the  anti-holomorphic 
involution $\hat \sigma$ induces a Pin-odd
transformation of ten-dimensional spinors, which has to be 
counterbalanced by the Pin-odd action induced by $P_3$
to ensure compatibility with the definite chirality of fermions in Type IIB
supergravity.
Note that the inclusion of $P_3$ 
is consistent with the interpretation of $x^3$
as the coordinate that parametrizes the interval
that decompactifies in the F-theory limit.
A more detailed study of the $\cO$ involution in the context
of string theory on toroidal orientifolds is desirable and is left for future investigation.\footnote{Note that 
the action of $\cO$ is reminiscent of non-standard orbifold actions recently considered in \cite{Grana:2012zn}.}

We conclude this section by analyzing the
up-lift of the 3d action for
K\"ahler and complex structure moduli. 
This will establish a match of the 3d Spin(7)
moduli with the Neumann scalars of a 
4d theory. Let us start with 
the  K\"ahler moduli.
In  Type IIB language these are given by
\beq
T_\alpha = \tfrac{1}{2!} \int_{Y_3} \omega_{ \alpha} \wedge J_{\rm b}^2
+ i \int_{Y_3} \omega_{\alpha} \wedge C_4 \ ,
\eeq
where now $\omega_\alpha$ and $J_{\rm b}$ are understood as 
$(1,1)$-forms on the double-cover $Y_3$ of the base $B_3$.
Recall the split introduced before \eqref{eq:base_split} of $H^2(B_3)$ that 
translates into a split of $H^2(Y_3)$ into positive and negative
subspaces under the action of $\hat \sigma^*$.
Note also that an expression of the form $\int_{Y_3} \lambda_6$
survives the $\hat \sigma$-projection only if $\lambda_6$
is negative under $\hat \sigma^*$. 
Using Table \ref{table:parities} one finds that $C_4$ has negative parity under $P_3 \, \hat \sigma$.
This implies that the 4d
moduli after the $\cO$ quotient should transform under 
$P_3$ as
\beq
  P_3\text{-even}:\ \Re \, T_{\alpha_-},\ \Im \, T_{\alpha_+}\ , \qquad  \qquad 
  P_3\text{-odd}:\ \Re \, T_{\alpha_+},\ \Im \, T_{\alpha_-}\ ,
\eeq
The $P_3$-even scalars match exactly with the 3d moduli
that survive the $\sigma$ quotient on the Calabi-Yau fourfold $Y_4$
on the M-theory side. 

Let us now turn to  complex structure moduli.
From a Type IIB perspective, those correspond to complex structure
moduli of the threefold $Y_3$, D7-brane moduli, and the axion-dilaton.
The action of the anti-holomorphic involution $\hat \sigma$
on $Y_3$ is such that
\beq \label{Omega30}
    \hat \sigma^* \Omega^{3,0} = e^{2i\theta} \, \overline{ \Omega^{3,0}}\ .
\eeq
This is completely analogous to the corresponding $\sigma$-action on the fourfold $Y_4$. 
Imposing \eqref{Omega30} one infers that the $P_3$-even 
complex structure moduli span a real subspace of the 4d $\cN=1$
moduli space.  With similar arguments it is possible to check the correspondence between 3d
Spin(7) moduli and 4d $P_3$-even moduli related to 
D7-branes and the axion-dilaton.

It is important to highlight the generic presence of $P_3$-odd 
scalars. Such scalar
degrees of freedom cannot have a constant non-vanishing profile
along the $x^3$ direction, and therefore do
not correspond to moduli in the 4d theory.
From a 4d perspective on a finite interval such scalars 
arise only as massive excitations.
In summary, we can state  that the orientifold picture suggests that 
the 4d  moduli, which are $P_3$-even, 
are in one-to-one correspondence with
the Spin(7) moduli in the 3d action \eqref{QuotientedY4action}.
The interpretation of the $P_3$-odd scalars from an 
M-theory perspective requires a better understanding 
of M2-brane states in our setup and seems not to be
accessible within the context of 11d supergravity.

\subsection{Aspects of Charged Matter}

The effective action derived in the previous sections does not furnish an explicit
description of the charged matter spectrum of F-theory on the class
of Spin(7) manifolds under consideration. 
Charged matter becomes massive after the gauge group is broken to the
Coulomb branch and is  integrated out. 

To get information about charged matter we can 
alternatively start looking at 
the weak coupling limit of our F-theory setup, discussed in the previous section.
It can contain D7-branes that wrap holomorphic cycles in the threefold $Y_3$ 
and have $(1,1)$-type 
world-volume flux to ensure the presence of 4d chiral fermions.
As we have seen, the  crucial new ingredient is  the anti-holomorphic involution $\hat \sigma$
combined with the transformation $P_3$ to have a symmetry of Type IIB.

We can specialize further and consider a point in moduli space
in which the Calabi-Yau threefold $Y_3$ is realized as a toroidal orbifold.
In this toroidal setups 
the embedding of D7-branes is described by one linear holomorphic
equation for the flat complex coordinates of the torus. 
Information about the charged matter spectrum can be obtained by first principles,
by quantizing open strings stretching between D7-branes.
We can make some general remarks on the interplay between
holomorphically embedded D7-branes and the anti-holomorphic involution.
First of all, the image branes are also holomorphically embedded,
if the anti-holomorphic action is linear in the flat coordinates of the torus.
Second of all, the world-volume flux of an image brane is still of $(1,1)$-type,
but its sign is reversed compared to the original brane.
These considerations imply that if we start with a supersymmetric setup
that contains only holomorphic branes with $(1,1)$ fluxes, these features 
are not spoiled by the introduction of image branes under the 
anti-holomorphic involution. 
Any intersection of any two branes or image branes possesses at least
one complex massless scalar.
Of course,
one has to take into account the projection onto invariant states to
determine if supersymmetry is actually present, or if different
number of bosonic and fermionic 
massless states is projected out.

It is possible to argue that the robust  features of the charged matter spectrum
are insensitive to the details of the full compactification setup, and only depend
on the local geometry around the intersection of the two D7-branes.
This can be effectively described by looking at a non-compact model with
flat D7-branes in  $\mathbb R^{1,3} \times \mathbb C^3$.
It captures the neighborhood of a fixed locus on the base $B_3$. Therefore
the anti-holomorphic action $\sigma$ in local coordinates 
can be taken to be one of the maps
given in \eqref{FixedPointsOnB}. If $\hat \sigma$ does not square to the identity, its square
is included as an additional holomorphic orbifold action, 
in such a way that $\hat \sigma^2 = \id$ in the quotient space.
We have performed explicitly the projection onto invariant states for 
the two linear actions in \eqref{FixedPointsOnB}, 
and we have compared the result with the purely orientifold
projection without the anti-holomorphic involution $\hat \sigma$
and without $P_3$. We have found 
that in both cases the same number of bosonic and fermionic degrees of freedom
survives the projection. This  signals that the charged matter spectrum is $\cN = 1$
supersymmetric also after the anti-holomorphic orbifold action is taken into account.

It can be checked that, irrespectively of the position of the D7-branes and their images
under the action of $\hat \sigma$,
no open string state can be invariant under the action of $\hat \sigma P_3$,
but rather that open string states are always swapped in pairs. This seems to prevent 
an undemocratic truncation of the spectrum in such a way that the same number 
of bosonic and fermionic degrees of freedom is obtained.
This general feature can be related to a mismatch between 
holomorphic embedding and anti-holomorphic involution.
On the one hand, charged matter is localized at the intersection of two D7-branes,
which is a complex one-dimensional holomorphic subspace of the internal 
six-torus. On the other hand,  the fixed locus of the 
anti-holomorphic involution is either a real one-dimensional subspace (see the first action in 
\eqref{FixedPointsOnB}), or a real
three-dimensional subspace incompatible with the holomorphic
structure (see the second action in \eqref{FixedPointsOnB}). 
It is therefore impossible to have the intersection inside the fixed locus of the 
anti-holomorphic involution.

There are many other interesting open questions that can be addressed in toroidal models. 
For instance, it might be possible to relate closed string twisted sectors of the 
anti-holomorphic orbifold action to resolution modes of the Spin(7) geometry.
We leave these investigations for future research.

\section{Conclusions}

In this paper we studied aspects of 4d effective theories arising from F-theory on 
Spin(7) manifolds. To approach this problem we proposed the duality \eqref{new_dual} between 
an M-theory compactification on a certain fibered Spin(7) geometry and F-theory
on a resolved version of this geometry multiplied by an interval,
in the shrinking fiber limit. This provides the opportunity to study 4d
theories from F-theory by using 3d minimally supersymmetric theories. 
We argued that these Spin(7) compactifications of F-theory can be approached via M-theory, 
when definite boundary conditions for the various 4d 
fields on the interval are chosen. Our analysis focused on the comparison 
of the 3d, $\cN=1$ zero mode actions on the M-theory and 
F-theory side. Up-lifting to four dimensions is the most conjectural 
step. However, making an appropriate minimal ansatz for the 4d theory \eqref{eq:4dactionansatz}
on a finite interval, its couplings 
and general features can be determined compatible with the Spin(7) reduction of M-theory. 
In particular one identifies the 4d scalar potential \eqref{gen4Dpotential} and fermionic couplings 
\eqref{Afunctions}. The study of the complete 4d Wilsonian effective action 
is complicated by the fact that in the F-theory limit M2-brane states will 
become light and can introduce new 4d degrees of freedom. These
can be light in the limit of a large interval and might help to restore 4d, $\cN=1$ 
supersymmetry away from the interval boundaries. This would be 
in the same spirit as \cite{Antoniadis:1999xk}.

To provide evidence for the M-theory to F-theory duality it was 
crucial to specify a class of Spin(7) manifolds for which this duality can be analyzed. 
Concretely, we employed Spin(7) manifolds that are obtained by quotienting an elliptically 
fibered Calabi-Yau fourfold by an anti-holomorphic and isometric involution $\sigma$. 
We called the resulting manifolds quotient torus fibered Spin(7) manifolds. 
Consequently, the 3d effective theories arising 
when compactifying M-theory on these quotient torus fibrations 
respect only the minimal number ($\cN=1$) of supersymmetries. These can 
be obtained by truncating the $\cN=2$ theory arising in the 
Calabi-Yau fourfold reduction of M-theory. 
More generally, we have revisited the 3d, $\cN=1$
effective theories arising in M-theory compactifications on 
smooth Spin(7) manifolds with four-form fluxes. We 
determined the characteristic data of the zero mode 
theories in terms of the internal geometry. 
The 3d theories obtained from the quotiented Calabi-Yau spaces 
were shown to comprise a special class of such 3d, $\cN=1$ theories.
The same class of theories was then shown to 
arise from an interval reduction of specific 4d non-supersymmetric 
theories if appropriate boundary conditions on the 4d fields are imposed.

The spectrum and couplings of the 4d theories 
were constrained to yield a 3d zero mode action that matches 
with the Spin(7) reduction. This imposes stringent constraints  
on the allowed 4d theories. We have argued that 4d vectors can still be grouped
with fermions similar to 4d, $\cN=1$ vector multiplets. The classical 
couplings of the vector fields and their fermionic partners  
are determined by 3d couplings of the 3d, $\cN=1$ 
Coulomb branch scalars $\phi^i$ 
by Lorentz symmetry and gauge symmetry. 
These symmetries should be restored in the interior of 
a very large interval. Similarly, one can proceed with the 
couplings of the 4d metric and gravitino that are constrained 
by the 3d couplings and 4d Lorentz symmetry. 
It should be stressed, however, that the 4d couplings are 
no longer supersymmetric in the minimal action \eqref{eq:4dactionansatz}
that can be unambiguously determined from the 3d M-theory reduction. 
This is due to the fact that one of the two real scalars in 
a 4d $\cN=1$ chiral multiplet would need to arise 
in the M-theory to F-theory limit from M2-brane states. 
In fact, we argued that one can map the Spin(7) moduli to 
only one of the real scalars in these multiplets. 
The kinetic terms of the 
real moduli scalars and the form of the scalar potential 
were discussed in section \ref{sec:4daction}. They are 
less constrained than in 4d, $\cN=1$ 
theories, but still inherit special properties from 
the class of Spin(7) geometries used in our work. Since 
for our construction there is always an underlying 
Calabi-Yau fourfold, 4d $\cN=1$ supersymmetry might 
be locally present away from the boundaries.
To complete this picture it would be desirable to study 
Kaluza-Klein modes arising in the interval compactification
and obtaining a 3d action including the 
dynamics of all such modes.\footnote{For actions of this 
type in a circle reduction, see, for example, \cite{Hohm:2006rj,Bonetti:2012fn}.}

In should be noted that the constructions of Spin(7)
manifolds originally proposed in \cite{Joyce:1999nk}
also included isolated orbifold points coinciding with the 
fixed  points of $\sigma$. Showing that these can be 
resolved in a Spin(7)-compatible fashion was a crucial task in \cite{Joyce:1999nk}. 
We have not included a study of these modes in this work, but it would 
be very interesting to understand how they modify the 4d effective 
theory. In particular, we found that if $\sigma$ has only 
isolated fixed points on $Y_4$  that the torus must be pinched over these points.  This suggests an interesting 
link between the gauge theory dynamics and the 
singularities that need to be resolved in a Spin(7)-compatible 
way to obtain a smooth geometry. As for ordinary 
non-Abelian gauge theory singularities of elliptically fibered 
Calabi-Yau fourfolds, F-theory might be well-defined on the singular Spin(7)
geometry if one can identify the new light states arising 
near the singularities. 

A complete understanding of the 
supersymmetry breaking in our proposed approach 
will require a more detailed understanding of the 
4d Wilsonian effective action. An important role in the
supersymmetry breaking is played by the presence of the boundaries. 
If one assumes a 4d Wilsonian action of the form \eqref{4dN=1boundary}
the scale at which the 4d $\cN=1$ supersymmetry is 
broken is related to the size of the inteval. 

A further interesting open problem is to understand 
which corrections the 4d action of scalars in our reduction will admit. 
In particular, the scalar fields $\phi^\alpha_{\rm b}$ and the 
volume $\cV_{\rm b}$ of the base $B_3$ are massless 
in the Spin(7) holonomy reductions presented here and 
it would be interesting to see which effects render these 
fields massive as, similarly to Calabi-Yau fourfold compactifications, $G_4$ fluxes cannot stabilize all K\"ahler moduli. We should, 
however, stress that we used fluxes only without including 
their back-reaction and it would be interesting to consider 
reductions on the back-reacted backgrounds of 
\cite{Becker:2000jc,Martelli:2003ki,Tsimpis:2005kj}.

Let us close by noting that many aspects of our proposal have 
only been addressed very briefly despite their imminent importance. 
We have only briefly discussed the weak coupling 
interpretation and the charged matter spectrum. 
The quantization of Type IIB string theory 
on the constructed backgrounds is an interesting open task 
to which we hope to return in the near future. This should also
shed more light on the string interpretation of the singularities 
induced in the Spin(7) construction and the presence of an interval.
The crucial observation has been that a simple circle  
reduction cannot connect the 4d and 3d effective
theories of F-theory and M-theory.   
This might admit alternative realizations, for example in  
Sherck-Schwarz reductions, which provide exciting 
further directions to implement such dualities.

\vspace*{.5cm}
\noindent
\subsection*{Acknowledgments}

We would like to thank 
Carlo Angelantonj, Jan Keitel, Dieter L\"ust, Wati Taylor, Paul Townsend, and Hagen Triendl for interesting discussions
and correspondence. We are particularly grateful to Eran Palti and Ralph Blumenhagen for their useful comments 
on the first version of this work. 
T.G.~would like to thank High Energy group of Harvard University for hospitality. 
This work was supported by a research grant of the 
Max Planck Society.

\begin{appendix}
\vspace{2cm} 
\noindent {\bf \LARGE Appendices}

\section{Conventions}
For every space-time dimension $d$
we choose the mostly plus signature for the metric $g_{\mu\nu}$
and we adopt the following conventions for 
the Riemann tensor:
\begin{align}
\Gamma^\rho_{\phantom{a}\mu\nu} &= \tfrac{1}{2} g^{\rho\sigma} \left( \partial_{\mu} g_{\nu\sigma} 
+  \partial_{\nu} g_{\mu\sigma} -  \partial_{\sigma} g_{\mu\nu}\right) \;,  \\
R^{\lambda}_{\phantom{a}\tau \mu\nu} &= \partial_{\mu} \Gamma^\lambda_{\phantom{a}\nu\tau} 
- \partial_{\nu} \Gamma^\lambda_{\phantom{a}\mu\tau}
 + \Gamma^\lambda_{\phantom{a}\mu\alpha} \Gamma^\alpha_{\phantom{a}\nu\tau} 
- \Gamma^\lambda_{\phantom{a}\nu\alpha} \Gamma^\alpha_{\phantom{a}\mu\tau} \;,
&
\nn  
R_{\mu\nu} &= R^{\lambda}_{\phantom{a}\mu\lambda\nu} \;, & R &= R_{\mu\nu} g^{\mu\nu} \;.
\end{align}
The Levi-Civita tensor is denoted by $\epsilon_{\mu_1 \dots \mu_d}$.
In our conventions it satisfies 
\begin{equation}
\epsilon_{01 \dots (d-1)} = \sqrt{-\det g_{\mu\nu}} 
\end{equation}
in any coordinate system $(x^0, x^1, \dots ,x^{d-1})$.
Differential $p$-forms are expanded on the basis of differential of the coordinates as
\begin{equation}
\lambda = \tfrac{1}{p!} \lambda_{\mu_1 \dots \mu_p} \; dx^{\mu_1} \wedge \dots \wedge dx^{\mu_p} \;,
\end{equation}
so that the wedge product of a $p$- and a $q$-form satisfies
\begin{equation}
(\alpha \wedge \beta)_{\mu_1 \dots \mu_{p+q}} 
= \tfrac{(p+q)!}{p!q!} \alpha_{[\mu_1 \dots \mu_p} \beta_{\mu_{p+1} \dots \mu_{p+q}]} \;.
\end{equation}
Exterior differentiation of a $p$-form is given by
\begin{equation}
(d\alpha)_{\mu_0 \dots \mu_p} = (p+1) \partial_{[\mu_0} \alpha_{\mu_1 \dots \mu_p]} \; .
\end{equation}
The Hodge dual of $p$-form in real coordinates and arbitrary space-time dimension $d$ 
is defined by the expression
\begin{equation}
(*\alpha)_{\mu_1 \dots \mu_{d-p}} = \tfrac{1}{p!} \alpha^{\nu_1 \dots \nu_p} 
\epsilon_{\nu_1 \dots \nu_p \mu_1 \dots \mu_{d-p}} \; ,
\end{equation}
in such a way that
\begin{equation}
\alpha \wedge *\beta = \tfrac{1}{p!} \alpha_{\mu_1 \dots \mu_p} \beta^{\mu_1 \dots \mu_p} \; *1
\end{equation}
holds for arbitrary $p$-forms $\alpha$, $\beta$.

\section{Example Spin(7) Holonomy Manifolds}

\subsection{A Hypersurface in a $\bP_{2,3,1}$ Fibration of $\bP_{1,1,1,1}$}

Let us consider a simple example of the construction described in Section \ref{Geoms} 
in which the Calabi-Yau fourfold $Y_4$ is described by a polynomial in a toric ambient 
space constructed by fibering the weighted projective space $\bP_{2,3,1}$ over 
$\bP_{1,1,1,1}$. In the language of toric geometry this is described by a reflexive 
polyhedron with the set of rays given in Table \ref{exampleData1}.
\begin{table}[h!]
  \centering
  \begin{tabular}{r@{\,$=$\,(\,}r@{,\;\;}r@{,\;\;}r@{,\;\;}r@{,\;\;}r@{\,)\;\;}|c|c@{\ \;\;}c@{\ \;\;}c@{\ \;\;}c@{\ \;\;}c@{\ \;\;}} 
   \multicolumn{6}{c|}{vertices} & coords. & $Q_1$ & $Q_2$ \\
    \hline\hline
     $\nu_1$ & $1$ & $0$ & $0$ & $0$ & $0$ &   	$x$$^\big.$ 	& $8$ & $2$ \\
     $\nu_2$ & $0$ & $1$ & $0$ & $0$ & $0$ &  	$y$  			& $12$ & $3$  \\
     $\nu_3$ & $-2$ & $-3$ & $0 $ & $0$ & $0$ & 	$z$  			& $0$ & $1$  \\
     $\nu_4$ & $-2$ & $-3$ & $-1$ & $-1$ & $-1$ & 	$u_1$  		& $1$ & $0$  \\
     $\nu_5$ & $-2$ & $-3$ & $1$ & $0$ & $0$ & 	$u_2$ 		& $1$ & $0$ \\
     $\nu_6$ & $-2$ & $-3$ & $0$ & $1$ & $0$& 	$u_3$ 		& $1$ & $0$ \\
     $\nu_7$ & $-2$ & $-3$ & $0$ & $0$ & $1$ & 	$u_4$  		& $1$ & $0$  \\
    \hline
  \end{tabular}
  \caption{\small Toric data for a reflexive polyhedron describing a $\bP_{2,3,1}$ 
  fibration of $\bP_{1,1,1,1}$.}
  \label{exampleData1}
\end{table}
\FloatBarrier
This gives a smooth ambient space in which the Calabi-Yau fourfold will be defined 
by a homogeneous degree $(24,6)$ polynomial in the $(Q_1,Q_2)$ identifications. 
This polynomial may be brought into the Weierstrass form \eqref{WeierForm} where 
now the coefficients $f$ and $g$ are degree $16$ and $24$, homogeneous polynomials 
of the base coordinates $\displaystyle{u_1,\ldots, u_4}$, respectively. A sufficiently 
general set of coefficients for these polynomials will then give a smooth Calabi-Yau 
fourfold. Next we impose a symmetry of this space under the action of the anti-holomorphic 
involution $\s$ where
\ba
\s ( u_1, u_2, u_3, u_4, x, y, z) &= ( \bar u_2, -  \bar u_1, \bar u_4, - \bar u_3, \bar x, \bar y, \bar z) \, . 
\label{identExamp1}
\ea
This restricts the coefficients of the polynomial. However these coefficients remain general 
enough that a generic polynomial is still non-singular. The identification $\s$ has no fixed 
space on the base, as the would-be fixed space $\displaystyle{u_1= u_2 = u_3 = u_4 = 0}$ 
is removed by the Stanley-Reisner ideal. Every point of the base then represents an example 
of situation \hyperref[situ1]{(1)} as described in section \ref{Geoms} and so the Spin(7) 
holonomy manifold\footnote{Note that 
strictly speaking the quotient manifold is expected to have ${\rm SU(4)}\times \bZ_2$ holonomy.} produced upon quotienting by $\s$ is non-singular. This means that no 
additional resolutions need to be performed. 

\subsection{A Complete Intersection in a $\bP_{1,1,1,1}$ Fibration of $\bP_{1,1,2,2}$} \label{sec:CIExample}

Next let us consider a second construction in which the ambient space is formed by 
fibering $\bP_{1,1,1,1}$ over $\bP_{1,1,2,2}$. In this case the Calabi-Yau is given by 
a complete intersection of two polynomials described  the following nef-partition
in Table \ref{exampleData2}.
\begin{table}[ht]
  \centering
  \begin{tabular}{c|r@{\,$=$\,(\,}r@{,\;\;}r@{,\;\;}r@{,\;\;}r@{,\;\;}r@{,\;\;}r@{\,)\;\;}|c|c@{\ \;\;}c@{\ \;\;}c@{\ \;\;}c@{\ \;\;}c@{\ \;\;}} 
    nef-part. &\multicolumn{7}{c|}{vertices} & coords. & $Q_1$ & $Q_2$ \\
    \hline\hline
    $\nabla_1$ & $\nu_1$ & $-1$  &   $-1$  &   $0$  &    $-1$ & $-2$ & $-2$ & $y_1$$^\big.$ & $1$ & $0$ \\
    & $\nu_2$ & $0$ & $0$ & $0$ & $1$ & $0$ & $0$ &$y_2$  & $1$ & $0$  \\
     & $\nu_3$ & $1$ & $0$ & $0$ & $0$ & $0$ & $0$ & $x_1$  & $1$ & $1$  \\
    & $\nu_4$ & $0$ & $1$ & $0$ & $0$ & $0$ & $0$  & $x_2$  & $1$ & $1$  \\

    \hline
    $\nabla_2$ & $\nu_5$ & $0$ & $0$ & $0$ & $0$ &  $1$ & $0$ & $v_1$$^\big.$ & $2$ & $0$ \\
    & $\nu_6$ & $0$ & $0$ & $0$ & $0$  & $0$ & $1$ & $v_2$$^\big.$ & $2$ & $0$ \\
    & $\nu_7$ &   $-1$  &   $-1$  &   $-1$  &    $0$ & $0$ & $0$ & $z_1$  & $0$ & $1$  \\
    & $\nu_8$ & $0$ & $0$ & $1$ & $0$ & $0$ & $0$ & $z_2$  & $0$ & $1$  \\
    \hline
  \end{tabular}
  \caption{\small Toric data for a nef-partition describing a C.I. in a $\bP_{1,1,1,1}$ fibration of $\bP_{1,1,2,2}$.}
  \label{exampleData2}
\end{table}
\FloatBarrier
The two polynomials $P_1$ and $P_2$ are then associated with the partitions 
$\na_1$ and $\na_2$ respectively. These are both degree (4,2) under identifications $(Q_1,Q_2)$.

In this case the base $\bP_{1,1,2,2}$ has a complex one-dimensional holomorphic 
orbifold singularity at $\displaystyle{y_1 = y_2 = 0}$ before considering any 
anti-holomorphic quotient. This lifts to two separate complex two-dimensional 
singular spaces in the total ambient space.  One, which is associated with 
the $Q_1$ identification, lies at $\displaystyle{y_1 = y_2 = x_1 = x_2 = 0}$ 
and the other, which is associated with the $ Q_1 - Q_2$ identification, 
lies at $\displaystyle{y_1 = y_2 = z_1 =z_2 = 0}$. 

Let us first consider the singular space which lies at $\displaystyle{y_1 = y_2 = x_1 = x_2 = 0}$. 
At this locus the polynomials can be written as
\ba
P_1 & = a_1 z_1^2 + b_1 z_1 z_2 + c_1 z_2^2 &
P_2 & = a_2 z_1^2 + b_2 z_1 z_2 + c_2 z_2^2
\ea
where $a_{1,2}$ $b_{1,2}$ and $c_{1,2}$ are homogeneous quadratics in $v_1$ and $v_2$. 
The singularities of the ambient space will then intersect both polynomials at the 
places where one of the roots of $P_1$ sits on top of one of the roots of $P_2$. 
At these points the \emph{resultant} of the pair of polynomials, given by
\ba
 -a_2 b_1 b_2 c_1 + a_1 b_2^2 c_1 + a_2^2 c_1^2 + a_2 b_1^2 c_2 - a_1 b_1 b_2 c_2 - 
 2 a_1 a_2 c_1 c_2 + a_1^2 c_2^2 \, , 
\ea
will vanish. This \emph{resultant} is a homogeneous octic in $v_{1,2}$ so gives 
eight $\bZ_2$ singular points on the Calabi-Yau fourfold at which the pair of the 
polynomials hit the two-dimensional space of singularities in the ambient space.

Next let us consider the singular space which lies at $\displaystyle{y_1 = y_2 = z_1 =z_2 = 0}$.
As before both polynomials will intersect the singularity of the ambient space 
when the resultant vanishes. This second resultant is a homogeneous quartic 
in $v_{1,2}$ so gives four $\bZ_2$ singular points. 

The Calabi-Yau fourfold may have extra singularities associated with the pinching 
of the torus. To find out where this happens we may make use of the singularity 
classification described in \cite{Esole:2011cn}. This shows that for a generic 
set of polynomial coefficients the torus pinches with a Type $I_1$ singularity 
over the intersection of a homogeneous degree $(72,0)$ polynomial in the 
$(Q_1,Q_2)$ identification, with the two polynomials that define the Calabi-Yau. 
Furthermore we find that this space intersects each of the $\bZ_2$ singular points described above.

We now impose a symmetry under the action of the anti-holomorphic involution $\s$ defined by,
\ba
 \s (y_1,y_2,v_1,v_2,x_1,x_2,z_1,z_2) = 
 (\bar y_2, - \bar y_1, \bar v_2, \bar v_1, \bar x_2, - \bar x_1, \bar z_2, \bar z_1) \ .
\ea
As before this constrains the coefficients of the polynomials but does not alter 
the singularity structure of the Calabi-Yau. We note also that in this case $\s$ 
is not an involution on its own but that the identification $Q_1$ must be used to make $\s^2 =\id$. 

The action of $\s$ on the base gives a real one-dimensional fixed line which 
sits inside the holomorphic orbifold singularity of $\bP_{1,1,2,2}$. At most 
places over this fixed line the torus is unpinched and has no fixed space. It  
represents an example of situation \hyperref[situ21]{(2.1)} described in 
Section \ref{Geoms}. However when the torus pinches over the fixed line of the 
base the pinched point on the torus becomes fixed under the action of $\s$ and 
so represents an example of situation \hyperref[situ3]{(3)}. In additional, these 
fixed pinched points on the torus also lie at the eight $\bZ_2$ singular points at 
$\displaystyle{y_1 = y_2 = x_1 = x_2 = 0}$. By comparison the four $\bZ_2$ 
singular points, which lie at $\displaystyle{y_1 = y_2 = z_1 =z_2 = 0}$ are 
not fixed under $\s$ but instead are mapped pairwise into each other.

The quotient of this Calabi-Yau by $\s$ then gives a singular Spin(7) manifold. 
The presence of these singularities is not a problem in F-theory as this is defined 
on singular spaces. However in order to use the M-theory duality we have 
described to find the effective action these singularities must be resolved in 
an appropriate fashion. It is unclear how one would carry out this resolution 
or even if such a resolution can be performed at all for this particular Spin(7) 
manifold. For this reason it will be extremely important to investigate these 
resolutions further in future work.

\end{appendix}



\end{document}